\documentclass[twocolumn, usenatbib]{mnras}
\usepackage{savesym}
\usepackage{graphicx}
\usepackage{longtable}
\usepackage{changepage}

\expandafter\let\csname equation*\endcsname\relax
  \expandafter\let\csname endequation*\endcsname\relax 
\usepackage{subfig}
\usepackage{amsmath}
\usepackage{amssymb}
\usepackage{verbatim}
\usepackage[yyyymmdd,hhmmss]{datetime}
\usepackage{array}
\usepackage{times}
\usepackage{xcolor}

\DeclareRobustCommand{\DE}[3]{#2}
\let\DEthebibliography\thebibliography
\def\thebibliography{\DeclareRobustCommand{\DE}[3]{##3}\DEthebibliography}

\newcommand\W {{W^r_{\ \phi}}}
\newcommand\code{{{\tt FitTeD}}}

\title[ Fitting transients with discs  ]{ Fitting transients with discs (\code): a public light curve and spectral fitting package based on evolving relativistic discs   }
\author [Andrew Mummery, et al. ]{Andrew Mummery$^1$\thanks{E-mail:
andrew.mummery@physics.ox.ac.uk},  Edward Nathan$^2$, Adam Ingram$^3$,  M Gardner$^2$\\
$^1$ Oxford Theoretical Physics, Beecroft Building,  Clarendon Laboratory, Parks Road, Oxford, OX1 3PU, United Kingdom \\
$^2$ Cahill Center for Astronomy and Astrophysics, California Institute of Technology, Pasadena, CA 91125, USA. \\
$^3$ School of Mathematics, Statistics and Physics, Newcastle University, Herschel Building, Newcastle upon Tyne, NE1 7RU, UK 
 }

\date{}

\begin{document}

\pagerange{\pageref{firstpage}--\pageref{lastpage}} \pubyear{2024}

\maketitle

\label{firstpage}

\begin{abstract} 
We present \code, a public light curve and spectral fitting Python-package based on evolving relativistic discs. At its heart this package uses the solutions of the time dependent general relativistic disc equations to compute multi-band light curves and spectra.  All relevant relativistic optics effects (Doppler and gravitational energy shifting, and gravitational lensing) are included. Additional, non-disc light curve and spectral components can be included to (for example) model the early time rise and decay of tidal disruption event light curves in optical-to-UV bands.  Monte Carlo Markov Chain fitting procedures are included which return posterior distributions of black hole and disc parameters, allowing for the future automated processing of the large populations of transient sources discovered by (e.g.,) the Vera Rubin Observatory. As an explicit example, in this paper we model the multi-wavelength light curves of the tidal disruption event AT2019dsg, finding a good fit to the data, a black hole mass consistent with galactic scaling relationships, and a late-time disc Eddington ratio consistent with the observed launching of an outflow observed in radio bands. 
\end{abstract}

\noindent

\section{Introduction}
Many transient astronomical sources derive their luminosity from an evolving accretion flow around a compact object, the emission from which can dominate observations taken across the electromagnetic spectrum. These multi-wavelength light curves in principle contain detailed information regarding the physics of the compact object at their heart, and offer an opportunity to probe the physical properties of astrophysical black holes.  The automated processing and analysing of time-domain astronomical data is a problem of increasing importance, as we enter the era of large all-sky surveys such as the Legacy Survey of Space and Time (LSST) at the Vera Rubin observatory, which is expected to discover thousands of tidal disruption events \citep[e.g.,][]{vanVelzen19b, Bricman20}. 

A wide range of astronomical sources, including Galactic X-ray binaries, so-called luminous fast-blue optical transients \citep[LFBOTs; e.g.,][]{Prentice18, Inkenhaag23}, and tidal disruption events \citep[TDEs; e.g.,][]{vanVelzen19, MumBalb20a, Wen20, Mummery_et_al_2024} are all thought to source (at least) some of their emission directly from evolving accretion flows. However, most models of accretion discs utilise steady state assumptions \citep[e.g.,][]{SS73, NovikovThorne73}, and the analysis of multi-epoch data typically treats each observational individually, with the physical parameters of the system allowed to vary in an unconstrained manner between observations. In reality, however, the physical parameters of the system are of course coupled between epochs, namely by the dynamical equations of motion which govern accretion flows (chiefly the constraints of mass and angular momentum conservation). As far as the authors are aware, no publicly available code currently exists which uses the self consistent evolutionary solutions of the relativistic disc equations to produce disc light curves, which can then be fit to data. 

In this paper we present such a Python-based public code: FITting TransiEnts with Discs (\code), which self consistently produces multi-band light curves and spectra from relativistic discs by solving the time-dependent equations of disc mass, angular momentum and energy conservation. All relevant relativistic optics effects (Doppler and gravitational energy shifting, and gravitational lensing) are included, meaning that all leading order effects of general relativity are included. Private, earlier, versions of this code have been used previously in the TDE \citep[e.g.,][]{MumBalb20a, Mummery_et_al_2024} and LFBOT  \citep{Inkenhaag23} literature. We envisage that this code will be of primary interest to the TDE community, but is more broadly applicable than to just TDEs and could for example be used to model soft-state light curve decays of X-ray binaries. 

This code differs markedly from the existing TDE light curve fitting model in {\tt MOSFiT} \citep{Guilochon18, Mockler19}, in that it infers the physical parameters of the system from the late-time optical/UV (and X-ray) data of a TDE, rather than the early time optical/UV emission, which {\tt MOSFiT} uses. 

The layout of this paper is the following. In section \ref{model} we introduce and discuss the physical content of the model. In section \ref{data} we discuss how \code\ handles astronomical data, and the statistical analysis framework we employ for fitting models to data. In section \ref{example} we present an example use case, and analyse the multi-wavelength lightcurves of the TDE AT2019dsg, before summarising the model, and potential systematics, in section \ref{summary}. 

\section{Physical model}\label{model}
In this section we introduce, in detail, the physics of each component of the \code\ models. The presentation here is likely more detailed than required for a typical \code\ user, but is intended to act as a reference. We begin by discussing the physics of the disc density evolution, before discussing how the local disc-frame emission is calculated. We then discuss the  physics of the photon trajectories and energy shifting  which determine the ultimately observed properties of the disc. Non-observable, but still interesting, quantities such as the bolometric luminosity and mass accretion rate are discussed, before we finish by summarising the non-disc emission components included in \code. These components will facilitate the modelling of multi-component systems, like the optical light curves of TDEs. 

\subsection{Disc density evolution}
The fundamental equation which \code\ solves describes the time evolution of the azimuthally-averaged, height-integrated disc surface density $\Sigma (r, t)$.    The disc is assumed to be sufficiently thin that its density profile is well described by a height-integrated form, and accreting at a sufficiently low luminosity (principally, a sub-Eddington luminosity) so that the effects of energy advection and pressure forces can be neglected. The disc is further assumed to be evolving in the mid-plane of the black hole. 

The central black hole is assumed to be described by the Kerr metric (i.e., the black hole is assumed to have zero electric charge).  The Kerr metric  describes the spacetime external to a black hole of mass $M_\bullet$ and angular momentum $J$.   Standard near equator cylindrical Boyer-Lindquist coordinates are used: $r$ (cylindrical radius), $\phi$ (azimuth), $z$ (height above equator), $t$ (time as observed at infinity), and $\text{d}\tau$ (invariant line element).   The ISCO radius, inside of which the disc is rotationally unstable, is denoted as $r_I$.  Other notation is standard: the gravitation radius is  $r_g = GM_\bullet/c^2$, and the black hole spin parameter (with dimensions of length) is $a = J/M_\bullet c$. We denote the dimensionless black hole spin $a_\bullet \equiv a/r_g$. The Kerr metric describes black holes whenever $|a_\bullet| \leq 1$. 

The disc fluid is assumed, to leading order, to follow circular orbits with a four-velocity denoted $U^\mu$ (related to coordinate $x^\mu$ by $U^\mu=\text{d}x^\mu/\text{d}\tau$).  The non-zero components of $U^\mu$ are the angular $U^\phi$ and temporal $U^0$ components. As is well known, an ionized plasma following circular orbits in a central gravitational potential is unstable to the megnetorotational instability \citep[or MRI;][]{BalbusHawley91} if a weak magnetic field is present, which there will be in an astrophysical setting. This instability causes the onset of disc turbulence, and drives angular momentum transport through the disc. 

We assume that there is an anomalous stress tensor present, $W^\mu_\nu$, due to this low-level disk turbulence.   The stress is a measure of the correlation between the fluctuations in $U^\mu$ and $U_\nu$ \citep[the contravariant counterpart to the 4-velocity][]{Balbus17}, and could also include correlated magnetic fields.  As its notation suggests, $W^\mu_\nu$ is a mixed tensor of rank two.    

The evolution equation for the disc surface density then follows from solving the coupled equations of rest mass and angular momentum conservation in the flow.  Under the above assumptions, the governing equation for the evolution of the disc may generally be written \citep{EardleyLightman75, Balbus17}:
\begin{equation}\label{fund}
{\partial \Sigma \over \partial  t} =  {1\over r U^0} {\partial \ \over \partial  r}\left({U^0\over U'_\phi}    {\partial   \over \partial  r} \left({r\Sigma \W \over U^0}\right)\right), 
\end{equation}
where the primed notation denotes a radial gradient. 
In the Newtonian limit (which corresponds to $r \gg r_g$, $U^0 \to 1$), this equation becomes the familiar \cite{LBP74} disc evolution equation \citep[e.g.,][]{Frank02}.

To solve this, and related, equations the functional forms of the various 4-velocity components of circular orbits in the equatorial plane of the Kerr metric are required, which we reproduce here for completeness
 \begin{align}
U_0 &= -{{\cal A} \over {\cal D}^{1/2}}, \quad 
  U^0 = {{\cal B} \over {\cal D}^{1/2}}, \quad 
 U_\phi = \sqrt{GM_\bullet r} {{\cal C} \over {\cal D}^{1/2}} , \nonumber \\ 
 &\quad U^\phi = \sqrt{{GM_\bullet}\over {r^3}} \, {1\over {\cal D}^{1/2}} , \quad  
  U_\phi ' ={1\over 2}\sqrt{GM_\bullet\over r} {{\cal BE} \over {\cal D}^{3/2}} , \nonumber \\  
 &\quad\quad\quad\quad\quad \Omega \equiv \frac{U^\phi}{U^0} = \sqrt{{GM_\bullet }\over {r^3}} {1\over {\cal B}} ,  \label{orb_eqs} 
 \end{align}
 where ${\cal A, B, C, D}$ \& ${\cal E}$ are all relativistic corrections which tend to 1 at large radii \citep[cf.][]{NovikovThorne73}
 \begin{align}
{\cal A} &= 1-2r_g/r +a\sqrt{r_g/r^3}, \\
{\cal B} &= 1+a\sqrt{{r_g}/{r^3}}, \\
{\cal C} &= 1 + {a^2}/{r^2} - 2a\sqrt{{r_g}/{r^3}}, \\
{\cal D} &=  1 - {3r_g}/{r} + 2a\sqrt{{r_g}/{r^3}}, \\
{\cal E} &= 1 - 6r_g/r - 3a^2/r^2 + 8a\sqrt{r_g/ r^3} . 
\end{align}
Note that ${\cal E}(r_I)=0$ defines the ISCO radius, the inner limit of validity of our model. Clearly, the inclusion of general relativistic effects substantially increases the algebraic complexity of the governing disc equation. Fortunately, analytical Greens function solutions of this equation are known, provided that the stress tensor $\W$ depends only on radius as a powerlaw 
\begin{equation}\label{stressdef}
    \W = w \left({2r\over r_I}\right)^\mu, 
\end{equation} 
to which we shall hereafter restrict ourselves. 

These analytical solutions were derived in \citet{Mummery23a}, under the assumption that all disc properties vanish at the ISCO $r_I$. These solutions represent the subsequent global evolution of an initial disc ring of total mass $M_d$ (and formally zero initial width), which is located at $r=r_0$ at an initial time $t = t_0$. 

For all times $t > t_0$ (and radii $r>r_I$) these analytical solutions take the functional form 
\begin{multline}\label{green_s}
G_\Sigma(r, t; r_0, t_0) = \\ {M_d \over 2\pi r_I^2 c_0}  \sqrt{x^{-\alpha} f_\alpha(x) \exp\left(-{1 \over x} \right) \left(1 - {2\over x}\right)^{5/2 - 3/4\alpha} }\\ 
 {x^{-3/4 - \mu} \over \tau} \exp\left({-f_\alpha(x)^2 - f_\alpha(x_0)^2 \over  4\tau} \right) I_{1\over 4\alpha} \left({ f_\alpha(x) f_\alpha(x_0) \over  2\tau}\right),
\end{multline}
where 
\begin{equation}
c_0 = x_0^{(1 + 14\mu)/8} \left(1 - {2\over x_0}\right)^{3/4 - 3/8\alpha} \left( f_\alpha(x_0)\exp\left({1\over x_0}\right) \right)^{-1/2} ,
\end{equation}
is a normalisation factor ensuring that $M_d$ is the total initial disc mass. 
These solutions are approximate, and represent asymptotic leading order solutions. \citet{Mummery23a} verified that these solutions reproduce full numerical solutions of the relativistic disc equations with an accuracy  at the $\mathcal{O}(1\%)$ level. In this expression, $I_\nu$ is the modified Bessel function of order $\nu$, and $\alpha$ is related to the stress index $\mu$ via $\alpha = (3-2\mu)/4$. The function $f_\alpha(x)$ is given by 
\begin{multline}
f_\alpha(x) = {x^\alpha \over 2 \alpha} \sqrt{1 - {2\over x}}\left[1  - {x^{ - 1} \over { (\alpha - 1)}} {}_2F_1\left(1, {3\over 2}-\alpha; 2-\alpha; {2\over x}\right) \right] \\ + {2^{\alpha - 2} \over \alpha (\alpha - 1)}\sqrt{\pi} {\Gamma(2-\alpha)  \over  \Gamma({3/ 2} - \alpha)} ,
\end{multline}
where $_2F_1(a, b; c; z)$ is the hypergeometric function, $\Gamma(z)$ is the gamma function, and 
\begin{equation}
x \equiv 2 r / r_I. 
\end{equation}
The variable $x_0 = 2r_0/r_I$ is the (normalised) initial location of the disc material. The time variable $\tau$ is given by 
\begin{equation}
\tau \equiv \sqrt{2 \over GMr_I} {w \over r_I} \left(1 - {r_I \over r_0}\right) \left(t -t _0\right),
\end{equation} 
where $t$ is measured in physical units. It is important to note that $\tau$ as defined here is {\it not} equal to the time in units of the viscous timescale at the initial radius $\tau \neq (t-t_0)/t_{\rm visc}(r_0)$. For the stress parameterisation used in this work the Newtonian definition of the viscous timescale (which we shall adopt here) is \citep[e.g.,][]{Pringle81}
\begin{equation}
    t_{\rm visc} \equiv {2 \over (3 - 2\mu)^2} {\sqrt{GM_\bullet r_0^3} \over  w}\left({r_I \over 2r_0}\right)^\mu, 
\end{equation}
where the final factor ensures that it is the amplitude of the turbulent stress at the initial radius $r_0$ which enters the evolutionary timescale. 

Once black hole parameters and the initial disc mass and radius is set, one can either specify $w$ or $t_{\rm visc}$ thereafter closing the full set of equations. In \code\ the ``viscous'' timescale $t_{\rm visc}$ is taken as the input parameter instead of $w$, owing to its more obvious physical interpretation. 

While clearly algebraically complex, these solutions are rapid to compute, and include all thin disc relativistic effects.  The disc density evolution is entirely specified by the two black hole parameters $M_\bullet$ (which simply scales all lengths and times in the system) and $a_\bullet$ (which sets the inner disc radius in units of $r_g$); the disc mass $M_d$ (which simply scales the amplitude of the disc density) and  viscous timescale $t_{\rm visc}$ (which simply scales the evolutionary timescale of the disc density); and the initial condition parameters $r_0$ (initial location) and $t_0$ (initial time). 

The default \code\ disc model takes $\mu = 0$. We have verified numerically that this choice does not noticeably affect the computed light curves of the disc system, and the resulting parameter inference. 

Finally, it is important to stress that at its heart this model assumes a smooth evolution of the disc, which is likely a good approximation {\it on timescales not substantially shorter than the viscous timescale of the disc}. This approximation is common to all analytical models of accretion flows, which are in reality  inherently turbulent systems. This turbulence will violate this smooth evolution assumption on timescale much shorter than the disc viscous timescale, and we stress that short-timescale fluctuations in the disc lightcurves will not be captured by this model. This will likely lead to formally large discrepancies between the model and short-timescale observations of  tidal disruption events X-ray light curves in particular, where turbulent disc fluctuations are exponentially enhanced \citep{MummeryTurner24}. We recommend the binning of observed light curves (if the data allows), to mitigate this inherent property of these systems. 

\subsection{Disc frame emission}
Naturally, one does not observe the evolving density of an accretion disc. The observed properties of a disc are instead determined  by the evolving temperature profile of the flow, which may be related to the disc surface density through the constraints of energy conservation. 

The dominant $r-\phi$ component of the turbulent stress tensor $\W$ serves both to transport angular momentum outward (governed by equation \ref{fund}), but also to extract the free energy of the disc shear, which is then thermalised and radiated from the disc surface.   In standard $\alpha$-disc modelling \citep{SS73}, which we follow here, both the the extraction and the dissipation are assumed to be local processes.
Under these assumptions, the disc surface temperature $T(r, t)$ is given by \citep{Balbus17}
\begin{equation}\label{temperature}
\sigma T^4 = \frac{3}{4}\sqrt{GM_\bullet \over r^{5}} \W \Sigma \,  { {\cal B} \over {\cal D}^{3/2}} .
\end{equation}
where $\sigma$ is the  Stefan-Boltzmann constant. 

Once a set of free parameters are specified for the disc (equation \ref{green_s}) the disc temperature profile is  thereafter specified at all  radii and at all future times (eq. \ref{temperature}). Despite the complicated looking expressions involved, the gross asymptotic properties of these solutions are simple, and correspond with Newtonian sensibilities. Combining expressions (\ref{stressdef}), (\ref{green_s}) and (\ref{temperature}) and taking $r_I \ll r \ll r_{\rm out}$, one finds to leading order 
\begin{equation}
    T^4 \sim r^{-3} \, t^{-n}, \quad r_I \ll r \ll r_{\rm out}, 
\end{equation}
where $n = (4 - 2\mu)/(3 - 2\mu)$ is the index at which the bolometric luminosity decays \citep[e.g.][]{Pringle81}. The classical $T \propto r^{-3/4}$ result holds to leading order for all stress parametrisations.  

With the temperature of the accretion flow specified, we further assume that the resulting emission is quasi-thermal. This is an excellent assumption in many disc systems, provided the flow is optically thick to scattering (such as is the case for a tidal disruption event disc, for example). By this we mean that the  specific intensity of the emission, {\it in the rest frame of the disc}, is given by
\begin{equation}
    I_\nu^{\rm disc}(\nu, T) = {1\over f_c^4} B(\nu, f_c T) = \frac{2h\nu^3}{c^2f_{c}^4 } \left[ \exp\left( \frac{h\nu}{k f_{c} T} \right) - 1\right]^{-1} ,
\end{equation}
where $B(\nu, T)$ is the planck function, and $f_c$ is the so-called colour correction factor, and the frequency $\nu$ is the frequency of the radiation in the rest frame of the disc. 

Before we discuss the energy shifting of photons over the course of their trajectory from the disc frame to the observer, we briefly discuss the physics of the colour correction factor. 

While the surface temperature of the disc $T(r,t)$ is given by the constraints of energy conservation (equation \ref{temperature}), it corresponds physically to the temperature of the disc surface at a height above the midplane where the optical depth of the disc equals 1.  It is important to note, however, that the disc's central temperature is generally much hotter, and is given by \citep[e.g.,][]{Frank02} 
\begin{equation}
T_c^4 = {3\over 8} \kappa \Sigma T^4 ,
\end{equation}
where for standard astrophysical parameters $\kappa \Sigma \gg 1$ \citep[e.g.,][]{SS73}, where $\kappa \simeq 0.4$ cm$^2$g$^{-1}$ is the electron scattering opacity. 

A distant observer will see a blackbody spectrum with temperature $T$ emitted from each disc annulus if and only if the liberated disc energy can be fully thermalised in the disc atmosphere. However, on their path through the disc atmosphere, photons can either be absorbed and re-emitted (thus totally thermalising their energy), or they can undergo elastic scattering. Elastic scattering however, by definition, does not change the energy of the photon, and so if this process dominates in the disc atmosphere, photons will be observed to have the “hotter” temperatures associated with the altitudes closer to the disc midplane, not the disc’s $\tau = 1$ surface. This modifies the disc spectrum by the time it leaves the disc atmosphere, an effect which is modelled with the colour-correction factor $f_c$.

In \code\ we use the \cite{Done12} temperature-dependent model of the colour correction factor. For the lowest disc temperatures, below a critical temperature $T = 3\times10^4 {\rm K}$,  Hydrogen is neutral and the Hydrogen absorption opacity is extremely  large. This results in the full thermalisation of the liberated disc energy, meaning that the emitted disc spectrum is well described by a pure blackbody function with temperature $T$, i.e., 
\begin{equation}\label{col3}
f_{c}(T) = 1, \quad T(r, t) < 3\times 10^4\,  {\rm K} . 
\end{equation}
As the temperature increases above $3 \times 10^4$ K, the colour correction factor begins to increase. This results from the growing  ionisation fraction's  of both Hydrogen and Helium, which acts to reduce the total disc absorption opacity. As a result the electron scattering opacity begins to dominate, more photons are scattered out of the disc atmosphere,  and the typical temperature of observed photons increases.  \citet{Done12} model the colour correction factor in this regime as 
\begin{equation}\label{col2}
f_{c}(T) =  \left(\frac{T}{3\times10^4 {\rm K}} \right)^{0.82}, \, \quad 3\times10^4 \, {\rm K} < T(r, t) < 1\times10^5 \, {\rm K} .
\end{equation}
It should be noted that the choice of temperature index and normalisation in this expression were set so that the colour-correction factor was equal to $1$ at $T = 3 \times 10^4$ K, and was continuous in joining onto the Compton-scattering regime discussed below, and was not determined by fundamental atomic physics. This parameterisation did however accurately reproduce the results of full radiative transfer simulations \citep{Done12}. 
For the highest disc temperatures $T > 1\times10^5 {\rm K}$, electron scattering completely dominates the absorption opacity and the colour correction factor begins to saturate, a result of Compton down-scattering in the disc atmosphere, to
\begin{equation}\label{col1}
f_{c}(T) =  \left(\frac{72\, {\rm keV}}{k_B T}\right)^{1/9}, \quad  T(r, t) > 1\times10^5 \, {\rm K}. 
\end{equation}
This saturation leads to a maximum value of $f_{c} \approx 2.7$.

\subsection{Photon physics -- observer frame emission}
By moving through the Kerr spacetime photons both change their energy (as determined by a local observer), and have trajectories which deviate from straight lines (thus changing the observed image plane structure of the source). As a photon field moves through the vacuum, the quantity $I_\nu/\nu^3$ is conserved \citep[e.g.][]{MTW}, meaning that the observed specific intensity is given by 
\begin{equation}
    I_\nu^{\rm obs} = \left({\nu_{\rm obs} \over \nu_{\rm disc}}\right)^3 I_\nu^{\rm disc}(\nu_{\rm disc}, T) \equiv g^3 \, I_\nu^{\rm disc}( \nu_{\rm obs}/g, T) . 
\end{equation}
This expression defines the energy shift factor $g$. The energy of a photon, with 4-momentum $p_\mu$, as observed by an observer with 4-velocity $U^\mu$, is given by
\begin{equation}
    E = - U^\mu p_\mu, 
\end{equation}
meaning that the energy shift of the disc photons over their trajectory is given by
\begin{equation}\label{gfac}
    g = {U^\mu_{\rm obs} \, p^{\rm obs}_\mu \over U^\mu_{\rm disc} \,  p^{\rm disc}_\mu} = \frac{1}{U^{0}} \left[ 1+ \frac{p_{\phi}}{p_0} \Omega \right]^{-1} ,
\end{equation}
where $U^0$ and $\Omega$ are evaluated in the disc frame (see equation \ref{orb_eqs}), and we have assumed that the distant observer is at rest. The 4-momentum components $p_\phi$ and $p_0$ are conveniently constants of motion in the Kerr metric, and so can be evaluated in the observers frame of reference. Cosmological energy-shifting can also be incorporated into this framework by taking the above expression and modifying it by a factor 
\begin{equation}
    g_{\rm total} = g \times {1 \over 1 + z},
\end{equation}
where $z$ is the cosmological redshift of the source. By default \code\ performs all calculations neglecting cosmological redshift (i.e., it sets $z=0$ by default), although this may be changed by the user.

What is actually observed from a disc system is the specific flux density $F_\nu$ of the disc radiation. This is given formally by
\begin{equation}
F_{\nu}(\nu_{\rm obs}) = \int I^{\rm obs}_\nu (\nu_{\rm obs} ) \, \text{d}\Theta_{{{\rm obs} }} .
\end{equation}
In this expression $\text{d}\Theta_{{{\rm obs} }}$ is the differential element of solid angle subtended on the observer's sky by each element of the disc. 

\subsubsection{The {\tt GR\_disc} model }

In a purely Newtonian Universe, where photons travel in straight lines and undergo no energy shifting, this expression simplifies significantly. When observed at an inclination angle $i$ (from the disc rotation axis), each disc annulus would subtend an element of solid angle 
\begin{equation}
    {{\rm d}\Theta_{\rm obs}} \to {2\pi r \cos(i) \, {\rm d}r \over D^2} .
\end{equation}
Therefore the spectral luminosity of the disc ($\nu L_\nu \equiv 4\pi D^2 \nu F_\nu$) in this limit is given by 
\begin{equation}\label{GR_disc}
    \nu L_\nu(t) =  \frac{16\pi^2 h \nu^4 \cos(i)}{c^2}\int_{\rm disc} {r\, f_c^{-4} \, \, {\rm d}r \over \exp\left( {h\nu}/{k f_c T} \right) - 1}  ,
\end{equation}
where the notation $\int_{\rm disc} (\dots) \, {\rm d}r$ indicates the emission is summed over the entire radial extent of the disc. Note that in this Newtonian limit the emission profile retains angular symmetry, and depends only on one (radial) dimension in the disc. 

For observations taken at photon energies below the peak disc temperature $h \nu \ll k T_{\rm max}$ the above expression is an excellent approximation to the full relativistic calculation of the disc spectrum. The reason for this is simple, these low-energy photons predominantly originate in the outer-most disc regions, where relativistic effects are of minimal importance. The light curve model defined by equation (\ref{GR_disc}) therefore represents an excellent model of optical-UV emission in many black hole accretion disc systems, and is referred to as the {\tt GR\_disc} model in \code.  

In addition to the spectral luminosity of the disc at a given frequency, it is often more convenient to compare the integrated luminosity across a broad band (for example the 0.3-10 keV X-ray bandpass of {\it Swift XRT}). This quantity is trivially given by 
\begin{equation}
    L_{\rm band}(t) = \int_{\nu_1}^{\nu_2} \nu L_\nu(t) \, {\rm d}\ln \nu ,
\end{equation}
where $\nu_1$ and $\nu_2$ denote the lower and upper frequencies of the band respectively. This quantity can be computed simply once equation (\ref{GR_disc}) has been solved. 

\subsubsection{The {\tt GR\_disc\_plus} model }

\begin{figure}
  \includegraphics[width=.5\textwidth]{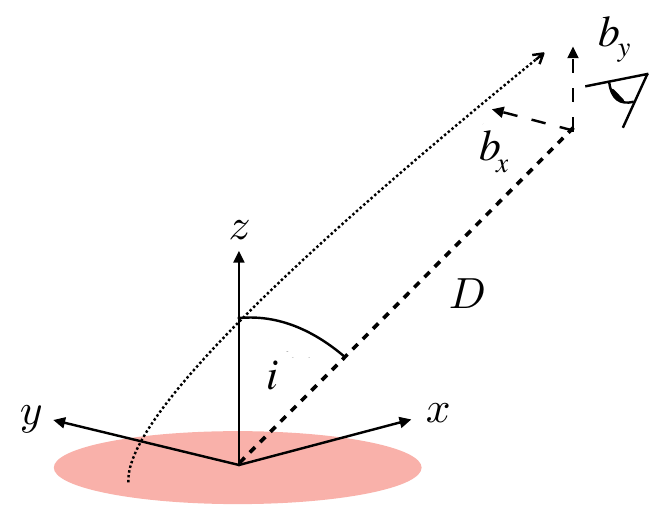} 
 \caption{Ray tracing geometry.  The coordinates $b_x$ and $b_y$ lie in the observer plane; $x$ and $y$  in the disc plane.   A schematic photon trajectory from the inner disc is shown.  The observer-disc inclination angle is denoted $i$.} 
 \label{RT}
\end{figure}

In a more relativistic Universe, however, computing the observed specific flux density is more complicated. For an observer at a large distance $D$ from the source, the differential solid angle in which the radiation is observed is
\begin{equation}
 \text{d}\Theta_{{{\rm obs} }} = \frac{\text{d}b_x \, \text{d} b_y}{D^2} ,
\end{equation} 
where $b_x$ and $b_y$ are the impact parameters at infinity (i.e., these are cartesian coordinates which describe the image plane of the telescopes ``camera'').  See {Fig.\  (\ref{RT})}  for further details.   
The observed spectral luminosity from the disc surface ${\cal S}$ is therefore formally given by
\begin{equation}\label{lum}
\nu L_\nu = 4\pi \nu _{\rm obs}  \iint_{\cal S} {g^3 f_{c}^{-4} B (\nu_{\rm obs} / g , f_{c} T)}\,  {\text{d}b_x  \text{d} b_y} .
\end{equation}
Note that $g$ depends upon $b_x$ and $b_y$ (through the 4-momentum of the observed photon, eq. \ref{gfac}).   There is no simple analytical solution which relates the quantity $g$ and the photon's emitted radius $r$ to the quantities $b_x$ and $b_y$ for general Kerr spin parameter $a$ and observer inclination $i$. This calculation must therefore be performed numerically. We summarise our algorithm below.

The distant observer is orientated at an inclination angle $i$ from the disc plane (at distance $D$; Fig. \ref{RT}). A photon at an image plane coordinate  $(b_x, b_y)$ has a corresponding spherical-polar coordinate $(r_i, \theta_i, \phi_i)$, given by \citep{Psaltis12}
\begin{align}
r_i &= \left(D^2 + b_x^2 + b_y^2\right)^{1/2}, \label{r0} \\
\cos\theta_i &=  r_i^{-1} \left( D \cos i + b_y\sin i\right) ,\\
\tan\phi_i &=   b_x \left(D\sin i - b_y\cos i \right)^{-1}  .
\end{align}
While in reality $D$ represents the astronomical distance to the source, provided $D\gg r_g$ all relativistic effects are included. \code\ uses $D = 10^4 r_g$, which includes all relativistic effects without compromising on computational speed.  The only photons which will contribute to the image have 3-momentum which is perpendicular to the image plane. This orthogonality condition uniquely specifies the initial photon 4-velocity \citep{Psaltis12}
\begin{align}
p^r_i &\equiv \left(\frac{\text{d} r}{\text{d}\tau'}\right)_{\text{obs}} = \frac{D}{r_i} , \\
p^\theta_i &\equiv \left(\frac{\text{d} \theta}{\text{d}\tau'}\right)_{\text{obs}}  = \frac{ D\left(D\cos i+ b_y\sin i \right) - r_i^2\cos i  }{r_i^2 \left( r_i^2 - \left(D \cos i   + b_y\sin i  \right)^2\right)^{1/2}} , \\
p^\phi_i &\equiv \left(\frac{\text{d} \phi}{\text{d}\tau'}\right)_{\text{obs}} = \frac{- b_x \sin i}{\left( D \sin  i - b_y\cos  i\right)^2 + b_x^2} , \label{up0}
\end{align}
from which the time component of the 4-velocity can be computed by solving $g_{\mu\nu} p^\mu p^\nu = 0$ in the observer plane. With the photon 4-velocity completely specified, the ratio $p_\phi/p_0$ can be computed in the observer plane. The final piece of information required to compute the flux contribution in each pixel is the radius at which the photon originated. To compute this we trace the rays back from the observer plane to the disc by solving the null-geodesics of the Kerr metric using the code \verb|YNOGK|, which is based on \verb|GEOKERR| \citep{YangWang13, DexterAgol09}. 

Starting from a finely spaced grid of points $(b_x, b_y)$ in the image plane, we trace the geodesics of each photon back to the disc plane \citep[see][for a detailed description of the image plane grid]{Ingram19}, recording the location at which the photon intercepts the disc plane $(r_d)$, and the ratio $p_\phi/p_0$ for each photon.  The parameter $r_d$ allows the disc temperature $T$ to be calculated at a given time $t$ (equation \ref{temperature}). The parameters $r_d$ and $p_\phi/p_0$ together uniquely define the energy-shift factor $g$. The integrand  (of eq. \ref{lum}) can therefore be calculated at every grid point in the image plane, and the integral (\ref{lum}) is then calculated numerically. 

As photons emitted in the outer disc undergo minimal energy shifting over their trajectory, and to a close approximation travel on straight lines, we only compute the spectral luminosity in this full numerical manner for the innermost  $300r_g$ of the disc. For the remainder of the disc we compute the spectral luminosity contribution using the Newtonian result (eq. \ref{GR_disc}). The total model spectral luminosity is given by the sum of the inner (eq. \ref{lum}) and outer (eq. \ref{GR_disc}) contributions. This model is called {\tt GR\_disc\_plus} in \code. Similarly to the {\tt GR\_disc} model, integrated luminosities across an observing band can also be computed.

\subsection{Other disc quantities}

\begin{figure*}
    \centering
    \includegraphics[width=0.45\linewidth]{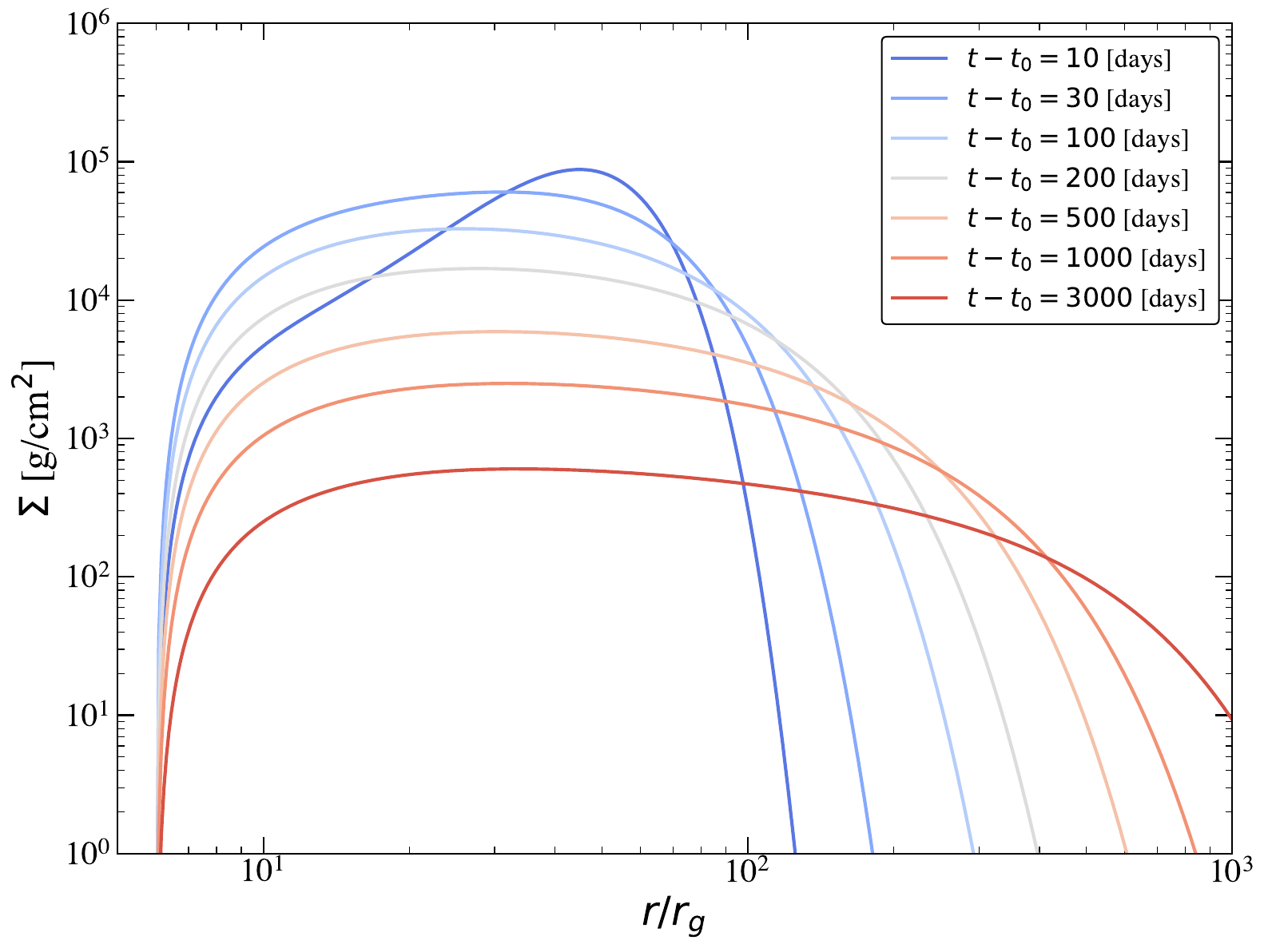}
    \includegraphics[width=0.45\linewidth]{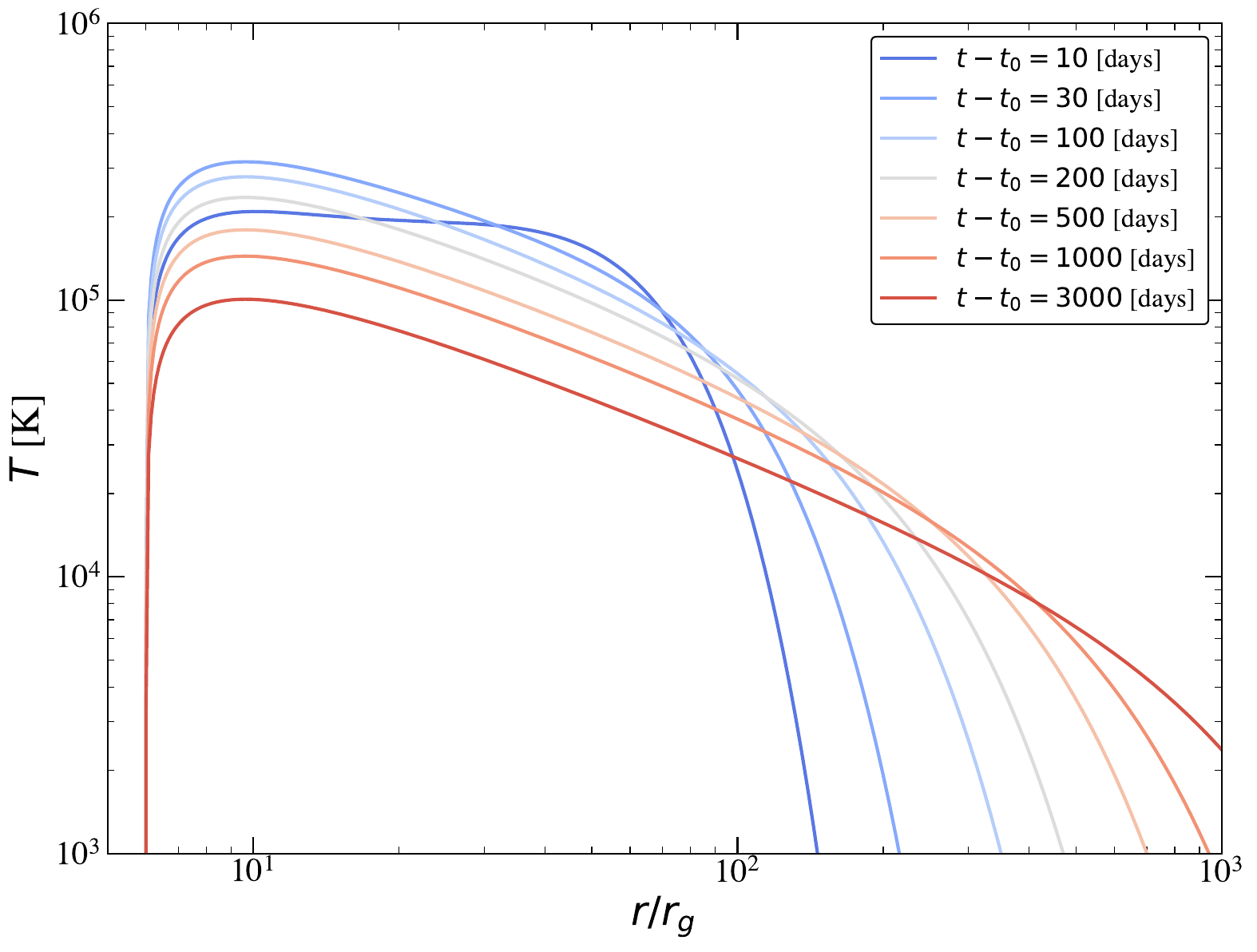}
    \includegraphics[width=0.45\linewidth]{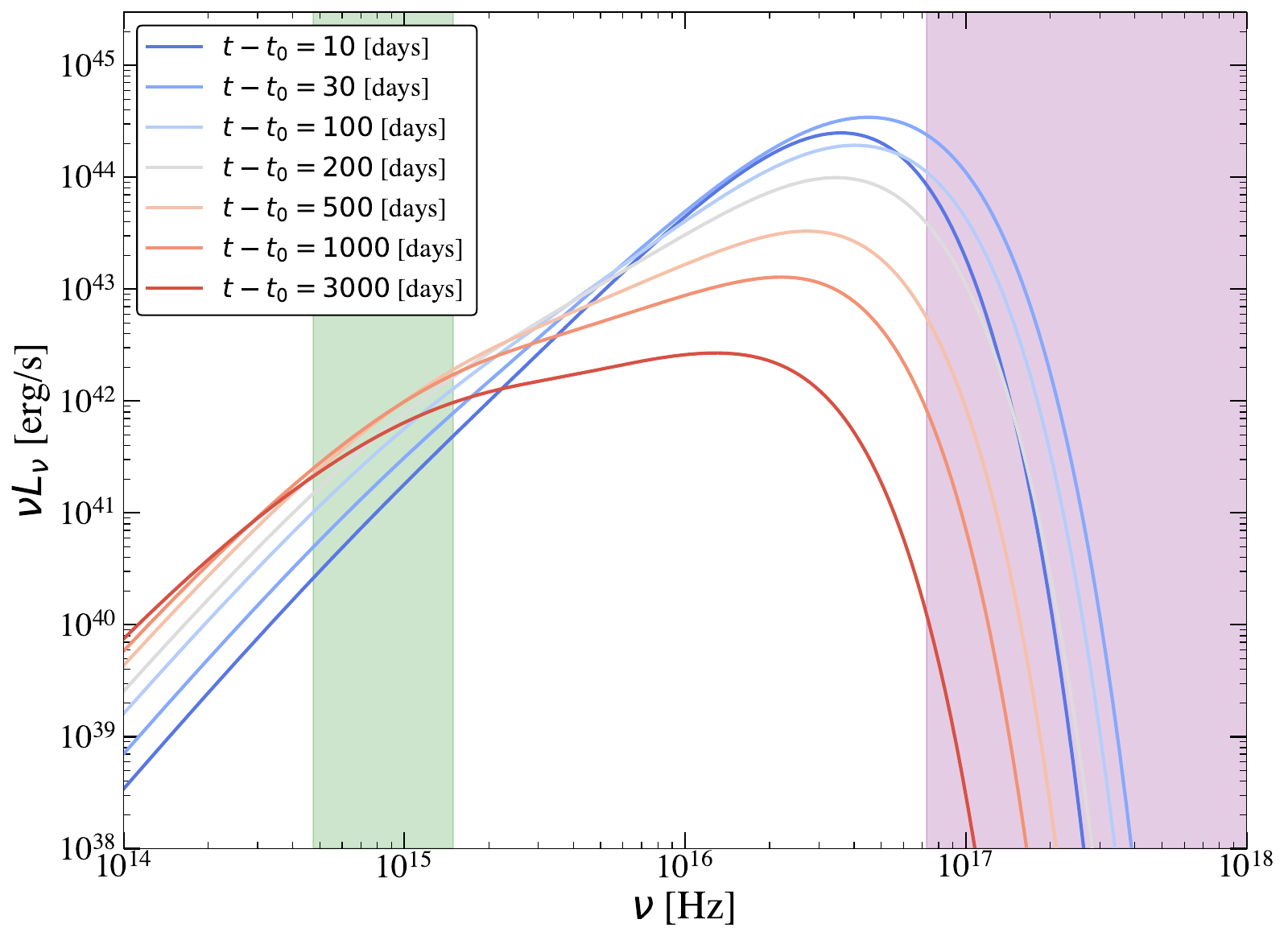}
    \includegraphics[width=0.45\linewidth]{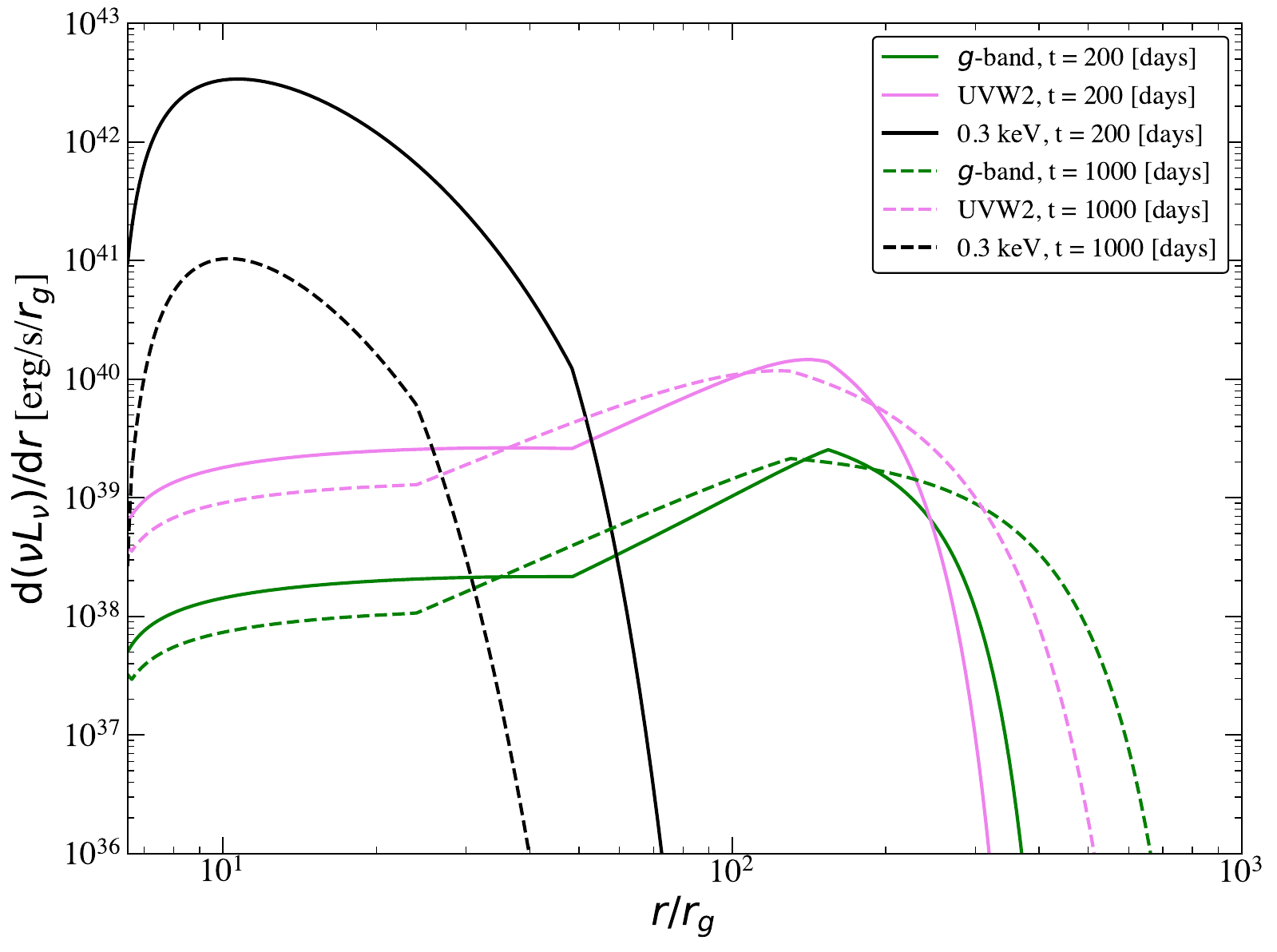}
    \includegraphics[width=0.45\linewidth]{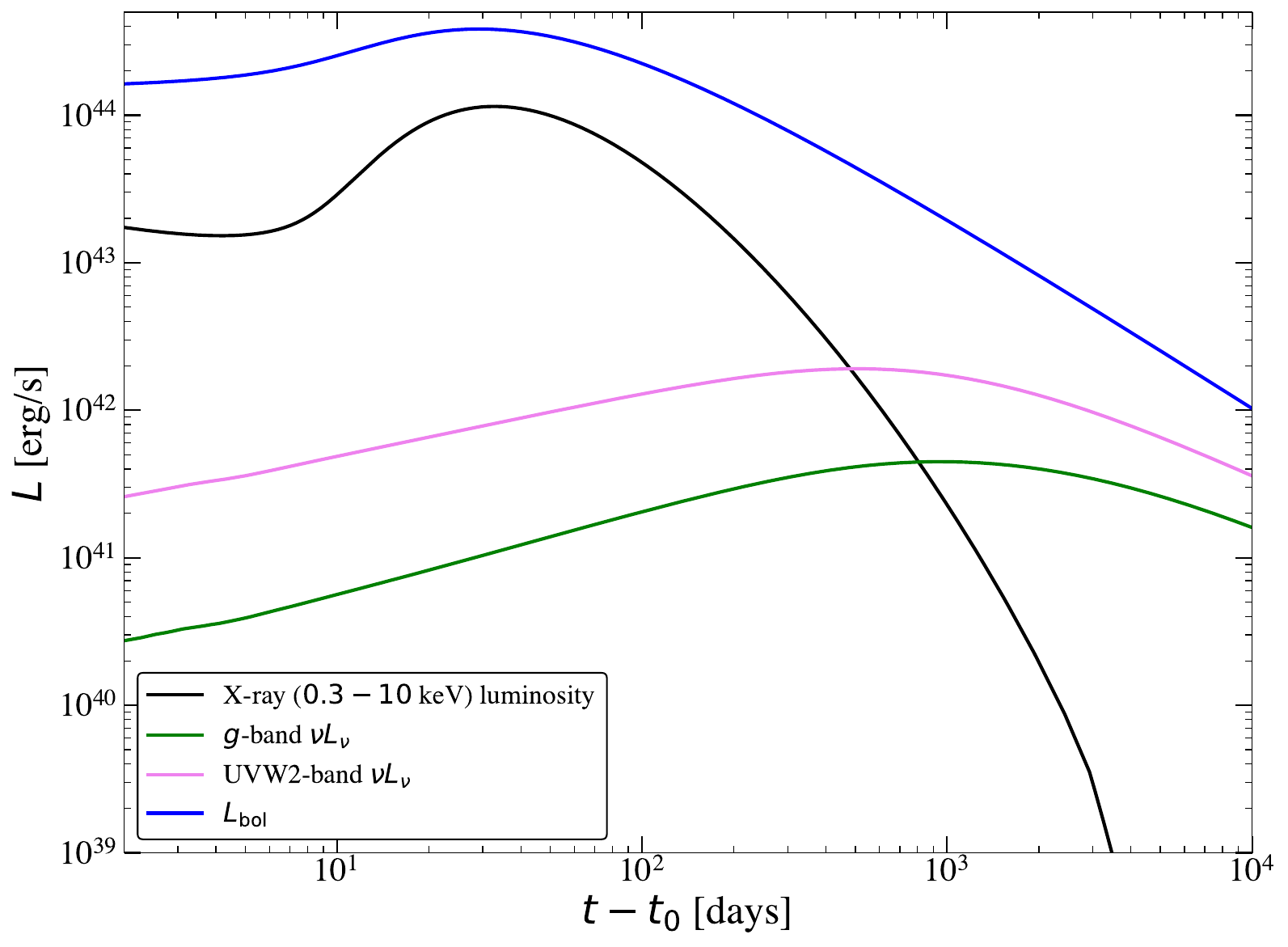}
    \includegraphics[width=0.45\linewidth]{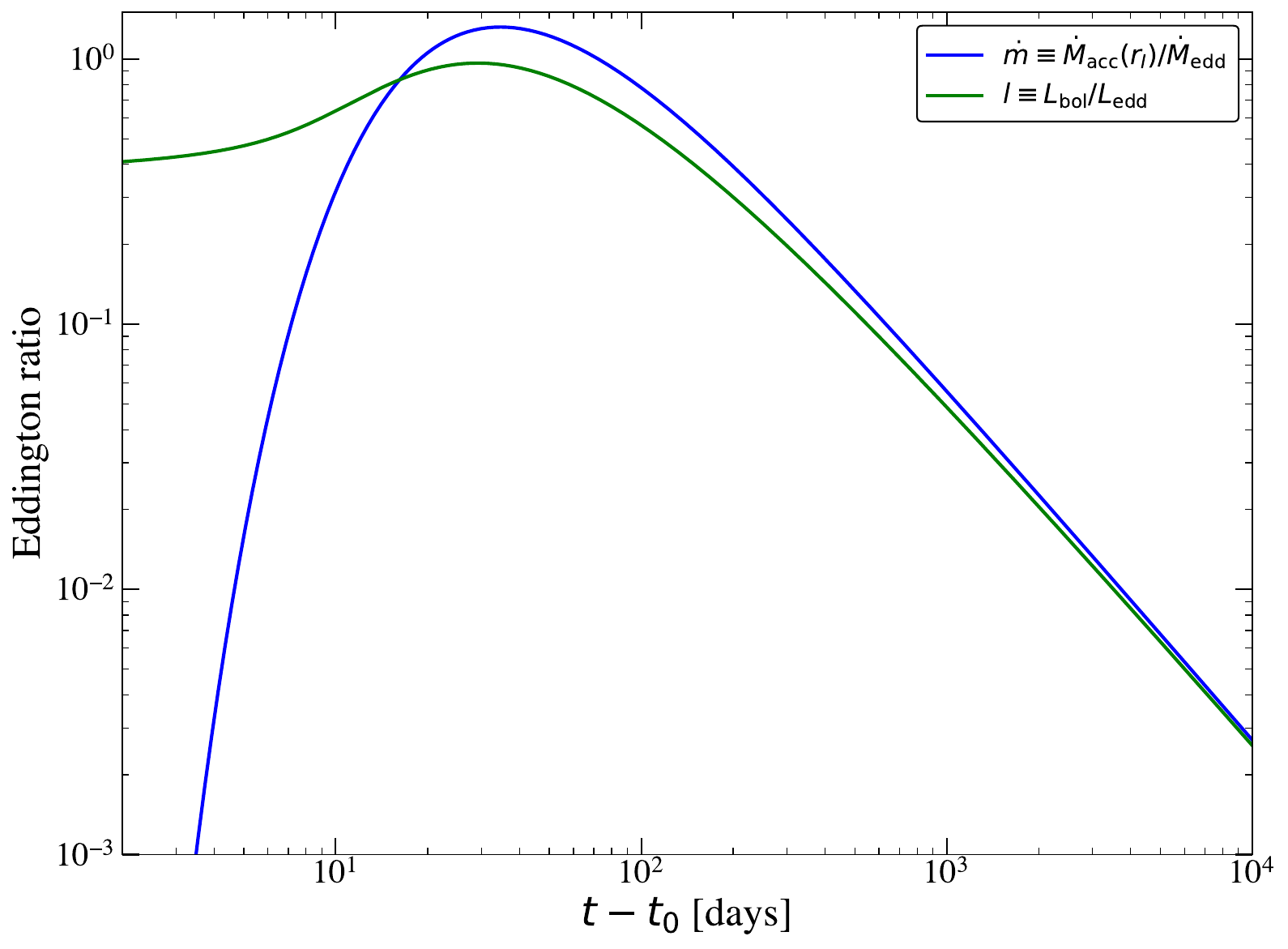}
    \caption{A set of example quantities which can be computed using the \code\ models {\tt GR\_disc} and {\tt GR\_disc\_plus}.  In the upper left panel we show the evolving disc density profiles, while in the upper right we show the corresponding temperature profiles. These temperature profiles produce the evolving disc spectra shown in the center left panel, where the green and purple vertical bands show typical optical/UV and X-ray observing bands respectively. Different disc locations provide different contributions to different spectral regions, as is highlighted in the center right panel. The light curves produced from this disc system, in the optical ZTF $g$-band, {\it Swift} UVW2 band and 0.3-10 keV X-ray band are shown in the lower left panel, along with the disc bolometric luminosity. The Eddington ratios (both luminosity and ISCO accretion rate) are shown in the lower right panel.   The parameters of this model were: $\log M_\bullet/M_\odot = 6.5$, $a_\bullet = 0$, $M_{\rm disc} = 0.1 M_\odot$, $r_0 = 50r_g$, $t_{\rm visc} = 100$ days, $t_0 = 0$ and $i = 45^\circ$, canonical TDE system parameters. }
    \label{fig:examples}
\end{figure*}

In addition to the spectral and band-integrated disc luminosity, there are a number of additional (although typically unobservable) quantities related to the disc which are of physical interest. 

The bolometric luminosity of the disc can be simply computed from the disc temperature as 
\begin{equation}
    L_{\rm bol}(t) = 2 \int_{\rm disc} 2\pi r \, \sigma T^4(r, t) \, {\rm d}r, 
\end{equation}
where the additional factor 2 here results from the two (upper and lower) surfaces of the disc which radiate. A relevant question for the validity of the \code\ disc models is whether this disc luminosity is sub-Eddington, where we define the Eddington luminosity as 
\begin{equation}
    L_{\rm edd} = {4\pi G M_\bullet m_p c \over \sigma_T} \simeq 1.26 \times 10^{38} \, \left({M_\bullet \over M_\odot}\right) \, {\rm erg}/{\rm s} , 
\end{equation}
where $m_p$ is the proton mass and $\sigma_T$ is the Thomson scattering cross section. The approximations governing the \code\ models will begin to break down when $L_{\rm bol} > L_{\rm edd}$, which should be checked on a source-by-source basis. 

Another interesting disc quantity is the disc's local mass accretion rate (a function of both radius and time), which is formally given by 
\begin{equation}
    \dot M(r, t) \equiv 2 \pi r U^r \Sigma, 
\end{equation}
and can be written explicitly in terms of the disc density and turbulent stress as \citep[e.g.,][]{EardleyLightman75, Balbus17}
\begin{equation}
    \dot M(r, t) = - {2\pi  U^0 \over U_\phi'} {\partial \over \partial r} \left({r \Sigma \W \over U^0}\right) , 
\end{equation}
where the prime denotes a radial gradient. Despite the angular momentum gradient of the flow vanishing at the ISCO (this is in fact the definition of the ISCO), this quantity converges to a finite limit as $r \to r_I$ \citep[and can be written explicitly for the Green's functions used here, see][]{Mummery23a}. 

Again, it is often of interest to compare the mass accretion rate (typically at the disc inner edge) with the Eddington mass accretion rate. In \code\ we define the Eddington mass accretion rate as 
\begin{equation}
    \dot M_{\rm edd} \equiv {L_{\rm edd} \over \eta c^2 }, 
\end{equation}
where 
\begin{equation}
    \eta(a_\bullet) = 1 - \left(1 - {2r_g\over 3r_I}\right)^{1/2}, 
\end{equation}
is the radiative efficiency of a vanishing ISCO stress black hole accretion flow as a function of black hole spin. Note that in a time dependent disc system $L_{\rm bol} \neq \eta \dot M(r_I) c^2$, although these two quantities do tend towards each other at large times as a quasi-steady state is approached.

A set of example quantities which can be computed using the \code\ models {\tt GR\_disc} and {\tt GR\_disc\_plus} are shown in Figure \ref{fig:examples}, for the model parameters $\log M_\bullet/M_\odot = 6.5$, $a_\bullet = 0$, $M_{\rm disc} = 0.1 M_\odot$, $r_0 = 50r_g$, $t_{\rm visc} = 100$ days, $t_0 = 0$ and $i = 45^\circ$. These parameters were chosen to represent canonical TDE system parameters.  In the upper left panel we show the evolving disc density profiles, while in the upper right we show the corresponding temperature profiles. These temperature profiles produce the evolving disc spectra shown in the center left panel, where the green and purple vertical bands show typical optical/UV and X-ray observing bands respectively. The optical/UV (green) band spans from the ZTF $r$-band to the {\it Swift} UVW2 band, while the X-ray (purple) band spans 0.3$-$10 keV.  

It is interesting to note that different disc locations provide different contributions to different spectral regions, as is highlighted in the center right panel, where we plot the quantity ${\rm d}(\nu L_\nu)/{\rm d}r$ (i.e., the integrand of equation \ref{GR_disc}). The X-ray luminosity (black curves) is dominated by the innermost disc regions, where the disc temperatures are highest. The optical/UV bands are however dominated by outer disc regions, where the effects of disc spreading offset the gradual disc cooling (keeping the lightcurves at these frequencies flat). The solid curves show the emission at a time near peak light ($t = 200$ days), while the dashed curves show a later time ($t = 1000$ days). 

The light curves produced from this disc system, in the optical ZTF $g$-band, {\it Swift} UVW2 band and 0.3-10 keV X-ray band are shown in the lower left panel, along with the disc bolometric luminosity. In this panel we see the exponential decay of the X-ray luminosity at late times \citep{MumBalb20a}, which contrasts with the power-law decline of the bolometric luminosity. The optical/UV light curves show only minor evolution over 10,000 days, as is typical for a TDE \citep{Mummery_et_al_2024}. The Eddington ratios (both luminosity and ISCO accretion rate) are shown in the lower right panel. This panel highlights how these quantities differ at early times, but converge at late times. 

\subsection{Non-disc rise and decay models}
While the models {\tt GR\_disc} and {\tt GR\_disc\_plus} account for the evolution of the transient disc emission, many astronomical sources are observed to have multiple superimposed emission components. As a relevant example, the early-time optical and UV observations of a tidal disruption event do not probe direct emission from the accretion flow, with this additional component thought to arise either due to reprocessing of disc emission \cite[e.g.,][]{Dai18} or emission from shocked gas heated in colliding debris streams \cite[e.g.,][]{Ryu20}. If these additional components are not satisfactorily modelled out, then systematic errors will effect the disc parameter inference. 

In \code\ we allow for the addition of early time components to model out any observed non-disc rise and/or decay, using the same methods as presented in \cite{Mummery_et_al_2024}. These models have the functional form  
\begin{equation}
    L_{\rm rise}(\nu, t) = L_0 \times f_{\rm rise}(t) \times \frac{\nu B(\nu, T)}{\nu_0 B(\nu_0, T)} , 
\end{equation}
and 
\begin{equation}
    L_{\rm decay}(\nu, t) = L_0 \times f_{\rm decay}(t) \times \frac{\nu B(\nu, T)}{\nu_0 B(\nu_0, T)} , 
\end{equation}
where $B(\nu, T)$ is the Planck function, and $\nu_0 = 6 \times 10^{14}$ Hz is a reference frequency (approximately equal to the ZTF $g$-band). The amplitude $L_0$ and temperature $T$ are common to both the rise and decay models. The functions $f_{\rm rise}(t)$ and $f_{\rm decay}(t)$ are defined so as to have a maximum amplitude of unity, so that $L_0$ remains the physical peak amplitude. 

We provide one rise and two decay models by default. The rise model is that of a Gaussian rise
\begin{equation}\label{gauss_rise}
    f_{\rm rise}^{\rm gauss}(t) = \exp\left( - {\left(t - t_{\rm peak}\right)^2 \over 2\sigma_{\rm rise}^2}\right) ,
\end{equation}
with fitting parameters $t_{\rm peak}$ and $\sigma_{\rm rise}$ (both with dimensions of time, and default units of days). 

The two decay models are that of an exponential decay
\begin{equation}\label{exp_decay}
    f_{\rm decay}^{\rm exp}(t) = \exp\left( - {\left(t - t_{\rm peak}\right) \over t_{\rm decay}}\right) ,
\end{equation}
with fitting parameters $t_{\rm peak}$ and $t_{\rm decay}$, and a power-law decay
\begin{equation}\label{pl_decay}
    f_{\rm decay}^{\rm pl}(t) = \left(  {t - t_{\rm peak} + t_{\rm fb} \over t_{\rm fb}}\right)^{-p} ,
\end{equation}
with fitting parameters $t_{\rm peak}$, $t_{\rm fb}$ and $p$. The parameters $t_{\rm peak}$, $t_{\rm fb}$ and $t_{\rm decay}$ have dimensions of time, and default units of days, while the index $p$ is dimensionless. The parameter $t_{\rm peak}$ is by default assumed to be common between the rise and decay models, and if no rise model is specified defaults to $t_{\rm peak} = 0$.  

In addition, the rise (decay) model is set to zero at times after (before) the fitted value for $t_{\rm peak}$. 

\subsection{Summary of models and parameters}
The parameters of the available  \code\ models  are summarised in Table \ref{tab_pars}. The physical premise of these models is that the observed transient emission of an astrophysical source is a superposition of direct disc emission (described by the {\tt GR\_disc} models), with a possible non-disc evolving thermal component. 
\begin{table*}
\def\arraystretch{1.5}
    \centering
    \begin{tabular}{cccc}
    \hline\hline
    Parameter name in \code\ & Physical meaning & Units & Default range  \\
    \hline
    {\tt log\_mh}  &  Black hole mass & $\log_{10} M_\odot $ & $(0, 10)$ \\ 
    {\tt a\_bh} &  Black hole spin parameter & Dimensionless & $(-1, +1)$ \\
    {\tt m\_disc} &  Initial mass of disc ring	& $M_\odot$ & $(10^{-3}, 10)$  \\
    {\tt r0} & Initial radius of disc ring & Gravitational radii  & $(r_I, 10^3)$ \\ 
    {\tt tvi} & Viscous timescale of disc evolution & Days & $(1, 10^4)$ \\ 
    {\tt t0} & Time of ring formation prior to $t = 0$ & Days & $(-100, 365)$ \\
    {\tt incl} & Disc-observer inclination angle & Degrees & $(0, 90)$ \\
    \hline\hline
    {\tt log\_L} & Amplitude of non-disc blackbody at frequency $\nu_0=6\times10^{14}$ Hz & $\log_{10}$ erg/s & $(0, \infty)$ \\
    {\tt log\_T} & Temperature of non-disc blackbody & $\log_{10}$ K & $(3.5, 5.5)$ \\
    \hline
    {\tt t\_decay} & Exponential decay timescale of non-disc emission & Days & $(1, 10^3)$ \\
    \hline
    {\tt t\_fb} & Powerlaw fallback timescale of non-disc emission & Days & $(1, 10^3)$ \\
    {\tt p} & Powerlaw decay index of non-disc emission & Dimensionless & $(0, 10)$ \\
    \hline
    {\tt t\_peak} & Time of peak of non-disc emission  & Days & $(-100, 100)$ \\
    {\tt sigma} & Gaussian rise timescale of non-disc emission & Days & $(1, 10^3)$ \\
    \hline\hline
     \end{tabular}
    \caption{A summary of the model parameters in the default  \code\ models. The ``Units'' column shows how to convert between \code\ units and physical units. The ``default range'' column shows the default bounds, in the format  (lower, upper), of the uniform window prior of each parameter.  }
    \label{tab_pars}
\end{table*}

The evolving disc emission is described by 7 free parameters. Two describe the central Kerr black hole: namely the mass $M_\bullet$ and spin $a_\bullet$ of the black hole. The disc itself is specified by four free parameters, the initial mass $M_{\rm disc}$ and radial location $r_0$ of the initial ring, which forms at a time $t_0$ before evolving with viscous timescale $t_{\rm visc}$. The final parameter specifies the angle between the observers line of sight and the black hole spin axis $i$. 

The initial condition used in the {\tt GR\_disc} models is highly simplified and somewhat arbitrary. It has the benefit of being  described by analytical solutions, which drastically reduces the computational expense of generating disc light curves, at the expense of some physical realism. Fortunately, solutions of diffusive equations like the governing relativistic disc evolution equation rapidly lose memory of their initial conditions, and approach self similar solutions at late times \citep[e.g.][]{Pringle81, Cannizzo90}. These self-similar solutions are precisely the Greens functions we use in this model, and the late time decay behaviour is typically most relevant for constraining physical parameters of the system. 

As this initial condition is somewhat arbitrary, we recommend (and provide by default) broad priors to be applied to the parameters $r_0$ and $t_0$. It is not immediately clear what the most realistic initial condition which should be applied in the case of an (e.g.,) tidal disruption event disc, and it is likely best not to over-constrain these parameters, at the detriment of the inference of more physically interesting parameters (such as the black hole mass, for example). 

In addition, we provide a variety of non-disc models of rising and decaying components, which are included to model out non-disc components in the observed spectrum of a source. These models are purely phenomenological, although they can be related to physical models for (e.g.,) early time TDE emission, and the best-fitting parameters may be of some interest for certain types of source. 

\section{Data treatment}\label{data}
\subsection{Data handling}
In \code\ a sources lightcurve data is separated by observing band, and is handled internally. The user uploads data in the format [times, luminosities, uncertainties, frequency], where the times, luminosities and uncertainties are lists of values recorded for each source, either at a given frequency or across a bandpass specified by a lower and upper pair of frequencies. These data are stored in a {\tt Data\_Set} class within \code. 

If the source to be modelled is one of the optical/UV tidal disruption events in the {\tt manyTDE}\footnote{{\url{https://github.com/sjoertvv/manyTDE}{}}} dataset \citep{Mummery_et_al_2024}, this data can be accessed directly by using the TDE's IAU name. 

We also include some simple data processing functions, including scripts which can remove sections of data (of use for perhaps the data taken prior to the peak of a TDE when no disc is present), and data re-binning scripts. Many disc systems are observed at cadences well below the viscous timescale of the disc, and on these timescales the luminosity is dominated by short timescale fluctuations which will not be captured by a smoothly-evolving model like those used in \code. We recommend the re-binning of lightcurve data for most optical/UV observations of tidal disruption events, for example. This has the added benefit of reducing the number of function evaluations required when fitting the \code\ models to light curve data. 

To be explicit, the re-binning splits the entire raw data set of a band into temporal bins of width $\Delta t_{\rm bin}$ (specified by the user), and then computes 
\begin{align}
    t_{{\rm new}, i} &= {1\over N_{{\rm bin}, i}} \sum_{{\rm data}, j \, \in \,  {\rm bin}, i} t_{{\rm old}, j} \\
    L_{{\rm new}, i} &= {1\over N_{{\rm bin}, i}} \sum_{{\rm data}, j \, \in \,  {\rm bin}, i} L_{{\rm old}, j} \\
    \delta L^2_{{\rm new}, i} &= {1\over N_{{\rm bin}, i}} \sum_{{\rm data}, j \, \in \,  {\rm bin}, i} \delta L^2_{{\rm old}, j} ,
\end{align}
where the summation is over all of the old data points within the $i^{\rm th}$ bin, and $N_{{\rm bin}, i}$ is the number of data points in each bin. Note that this process then reduces the average uncertainty of the new data points by $1/\sqrt{N_{\rm bin}}$. 

\subsection{Data fitting}
The fitting of models to data is done using standard maximum likelihood estimation techniques. 

For each set of parameters $\left\{ \Theta \right\} = [M_\bullet, a_\bullet, M_{\rm disc}, ...]$, we compute the total log-likelihood assuming each measurement is chi-squared distributed 
\begin{equation}
    \ln {\cal L}(\Theta) = - {1 \over 2}\sum_{{\rm bands}, \, i} \sum_{{\rm data, \, j}} \frac{\left(O_{i, j} - M_{i, j}\right)^2  }{E_{i, j}^2 + S_{i, j}^2} , 
\end{equation}
where $O_{i, j}$, $M_{i, j}$ and $E_{i, j}$ are the observed luminosity, model luminosity and luminosity uncertainty of the $j^{\rm th}$ data point in the $i^{\rm th}$ band respectively. The quantity $S_{i, j}$  here is a systematic error for a given data point, which may be optionally included by the user. This may be of use, for example, in re-weighting the relative importance of early/late time emission in optical/UV tidal disruption event models. Note that this definition neglects a contribution of $1/2 \ln 2\pi E_{i, j}^2$ for every data point, but as the purpose of this likelihood is to distinguish between different models fit to the same data, and this factor is common between all models, this does not effect the values of any best fitting parameters, only the absolute value of the likelihood of this fit (which is not of intrinsic interest). 

In addition, it is sometimes the case that no observable emission is detected from a transient source, where only upper limits on the source luminosity can be placed. This is particularly relevant for observations taken in the X-ray bands of tidal disruption events, many of which are found to be X-ray dim \citep[e.g.][]{Guolo24}.  These upper limits still contain important statistical information regarding the source, and can be included in the \code\ analysis. The following log-likelihood is used for upper limits 
\begin{equation}
    \ln {\cal L}_{U}(\Theta) = - {1 \over 2}\sum_{{\rm bands}, \, i} \sum_{\substack{{\rm limits, \, j, } \\ {\, M_{i, j}>U_{i, j}}}} \frac{\left(U_{i, j} - M_{i, j}\right)^2  }{(U_{i, j}/N_{\sigma, i, j})^2} , 
\end{equation}
where $U_{i, j}$ is the upper limit luminosity, with statistical significance of $N_{\sigma, i, j}$ standard deviations. The model luminosity is again $M_{i, j}$, and only those model values with luminosities greater than the upper limit contribute to the likelihood. 

As discussed in Table \ref{tab_pars}, the values of each free parameter are restricted, on physical grounds,  to lie between prior bounds which can be set by the user. We model these prior bounds with the following function 
\begin{equation}
    \ln {\cal P}(\Theta) = \sum_{{\rm parameters}, k} \ln \, \left[{\cal U}({\rm low}_k, {\rm high}_k, k) \right],
\end{equation}
where ${\cal U}(a, b, x)$ is the uniform distribution, which is equal to 1 if $x$ lies between ``$a$'' and ``$b$'',  and  equal to zero elsewhere. The default prior bounds for each parameter are summarised in Table \ref{tab_pars}. 

A best-fitting set of parameters is given by the parameter set $\{ \Theta_\star\}$ which maximises the probability of the model given the data, i.e., 
\begin{equation}
    \Theta_\star = \max_{\theta} \left[ \ln {\cal L}(\theta) +  \ln {\cal L}_U(\theta) + \ln {\cal P}(\theta)\right], 
\end{equation}
and is known as a maximum likelihood estimate (MLE) of the best-fitting parameters. The \code\ command {\tt best\_fit} computes the MLE parameters for a given data set. MLE methods are however known to be sensitive to small changes in observed data properties, particularly for relatively complex models with large parameter ranges like the {\tt GR\_disc} models. A more robust approach to determining best-fitting parameter values is to use Monte Carlo Markov Chain (MCMC) techniques. This can be performed using the {\tt run\_chain} command, which utilises the {\tt emcee} package \cite{EMCEE}. 

\section{Example --  tidal disruption event AT2019dsg}\label{example}

\begin{figure}
    \centering
    \includegraphics[width=\linewidth]{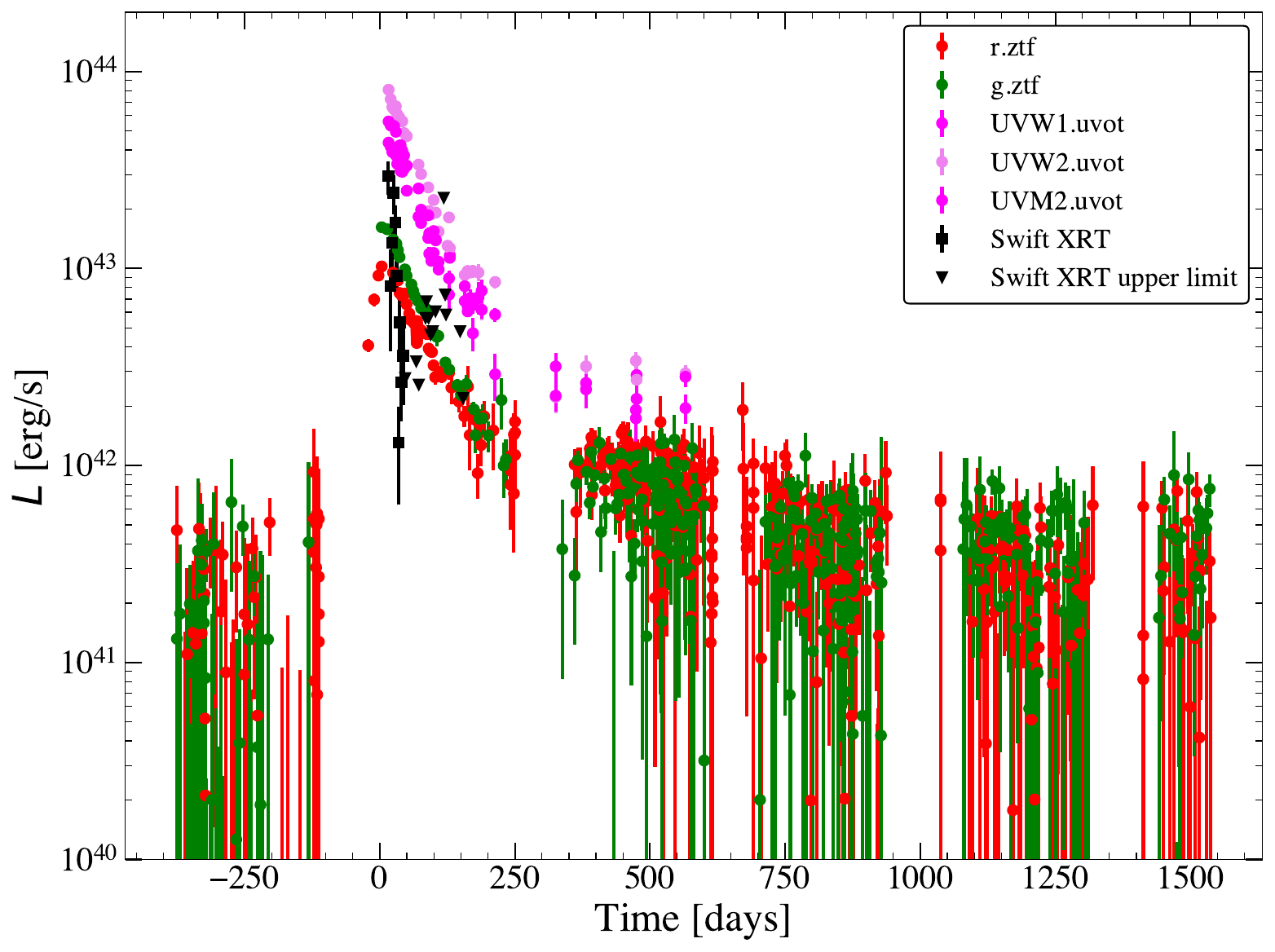}
    \includegraphics[width=\linewidth]{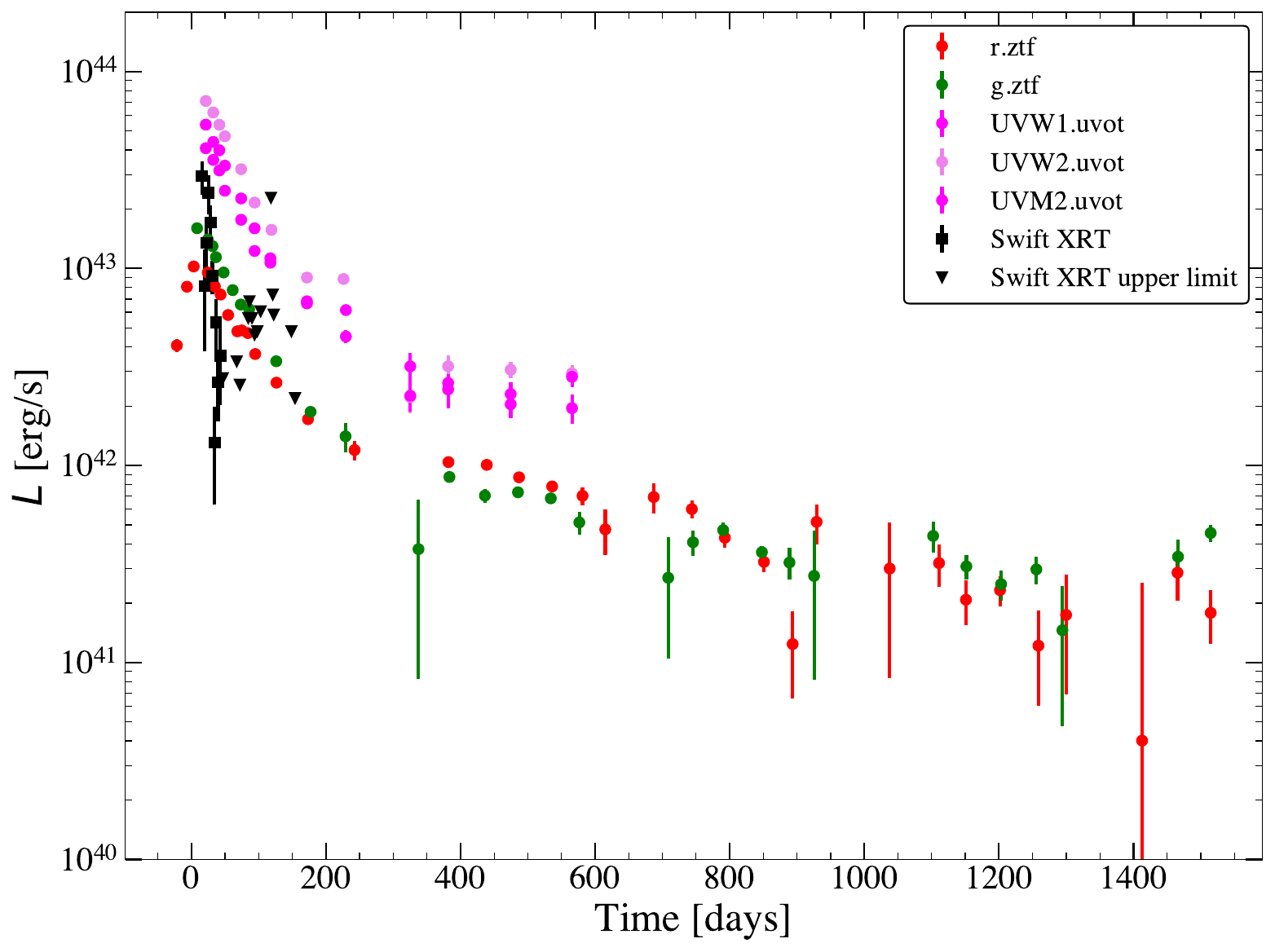}
    \caption{Upper: the raw optical--UV--X-ray light curve data of the tidal disruption event AT2019dsg. Lower: the light curves of AT2019dsg after re-binning the optical/UV emission, and neglecting the data prior to $t = -50$ days before peak flux. This figure highlights how ZTF data is widely available for TDEs, but is significantly noisier than data taken in ultraviolet bands. The X-ray emission at early times is a mixture of detections (black circles) and upper limits (black triangles).  }
    \label{fig:data_19dsg}
\end{figure}

As an example of using \code\ to constrain the physical parameters of a transient disc system, we analyse the tidal disruption event AT2019dsg \citep{Stein21, vanVelzen21}. This tidal disruption event was chosen as it has optical/UV lightcurves which are representative of a ``standard'' TDE, while also having both X-ray detections and non-detections at early times, therefore utilising the full \code\ likelihood. 
\subsection{Data processing}
In Figure \ref{fig:data_19dsg} we display both the full raw  data \citep[optical/UV data is taken from the {\tt manyTDE} data set][while the X-ray data is taken from \citealt{Stein21}]{Mummery_et_al_2024}, and the processed data set which we will use for lightcurve fitting. As can be clearly seen, AT2019dsg highlights the different data sets typically available to TDE science. There is a wealth of ZTF data in the $r$ and $g$ bands, but this is typically of lower quality, with each late-time observation (those observations most important for disc modelling) close to the detection threshold. In the {\it Swift UVOT} ultraviolet bands however, the data is much cleaner, but fewer observations are typically available. The light curve morphology is the same in both optical and UV bands (a rapid decay followed by a plateau). In X-ray bands AT2019dsg shows a rapid decay (quicker even than the optical/UV flux), followed by a series of non-detections.  The optical/UV data represents $\nu L_\nu$ for each band, and we take $\nu$ to be the effective frequency of each band. The X-ray data on the other hand is the integrated luminosity across the broad 0.3$-$10 keV {\it Swift XRT} bandpass.

We process the data in the following manner. We leave the X-ray data in its raw form, owing to the fewer data points collected at these energies.  In the optical/UV we first remove all data taken more than 50 days prior to peak light, as this data is just day-to-day noise with average zero flux. So as to well resolve the initial lightcurve decay (for fitting of the powerlaw model), we bin the first 100 days post-peak of optical/UV emission into 10 day bins, with the remaining optical/UV data binned on a 50 day timescale. The processed data can be seen in the lower panel of Figure \ref{fig:data_19dsg}. 

To aid in model fitting we added a global systematic of 10$\%$ of the luminosity in each band. We found that the $r$-band data was particularly noisy, and close to the detection threshold, so added a further 30$\%$ luminosity systematic error to only this band. 

\subsection{Model fitting and results}
We model the data in the lower panel of Figure \ref{fig:data_19dsg} with the \code\ model {\tt GR\_disc} + {\tt gauss\_rise} + {\tt pl\_decay}, where the rise/decay model is only for the optical/UV bands; in the X-ray bands we include no initial rise or decay components, as they are not needed physically (or statistically). 
\begin{figure}
    \centering
    \includegraphics[width=\linewidth]{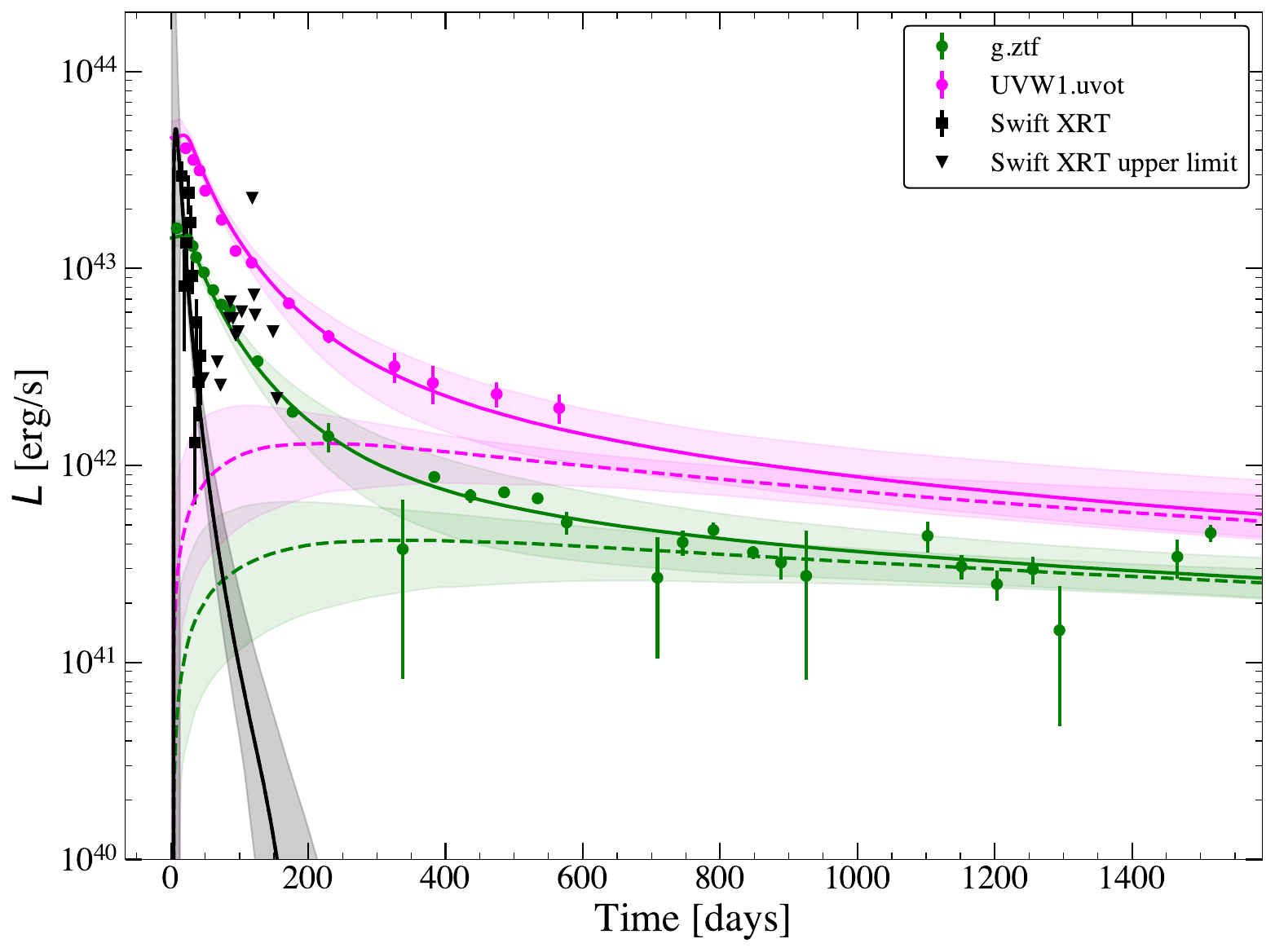}
    \includegraphics[width=\linewidth]{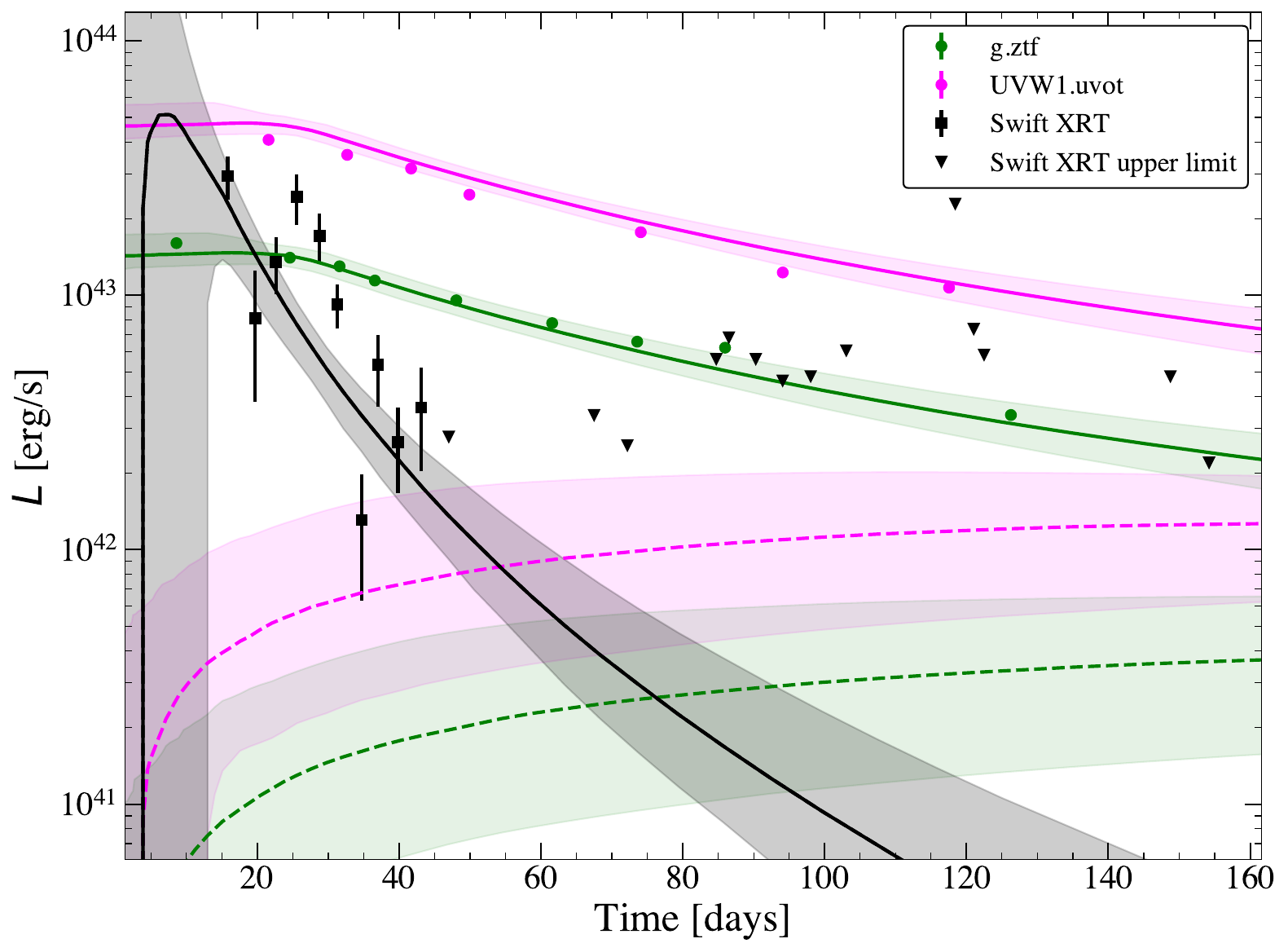}
    \caption{The lightcurve posterior results of a MCMC fit of the model {\tt GR\_disc} + {\tt gauss\_rise} + {\tt pl\_decay} to the AT2019dsg data. Only the X-ray, UVW1 and $g$-band data and posteriors are shown for clarity, but all of the processed data from Figure \ref{fig:data_19dsg} is used for fitting purposes. The dotted curves show the disc contribution to the combined disc and early time decay model (shown by solid curves) in optical and UV frequencies. There is no additional component required in the X-ray bands.  Shaded regions show the $5-95\%$ posterior range around the median (denoted by solid curves).  }
    \label{fig:GRdisc_example}
\end{figure}

We employ the default uniform priors on each parameter, as displayed in Table \ref{tab_pars}. We ran a 100 walker MCMC chain for 3000 steps  using {\tt emcee}, initialised from the MLE of the best-fitting parameters. Running this chain on a 2021 Apple MacBook Pro took roughly 0.1 seconds per walker per step (for a total time of around 7.5 hours). 

\begin{figure*}
    \centering
    \includegraphics[width=0.8\linewidth]{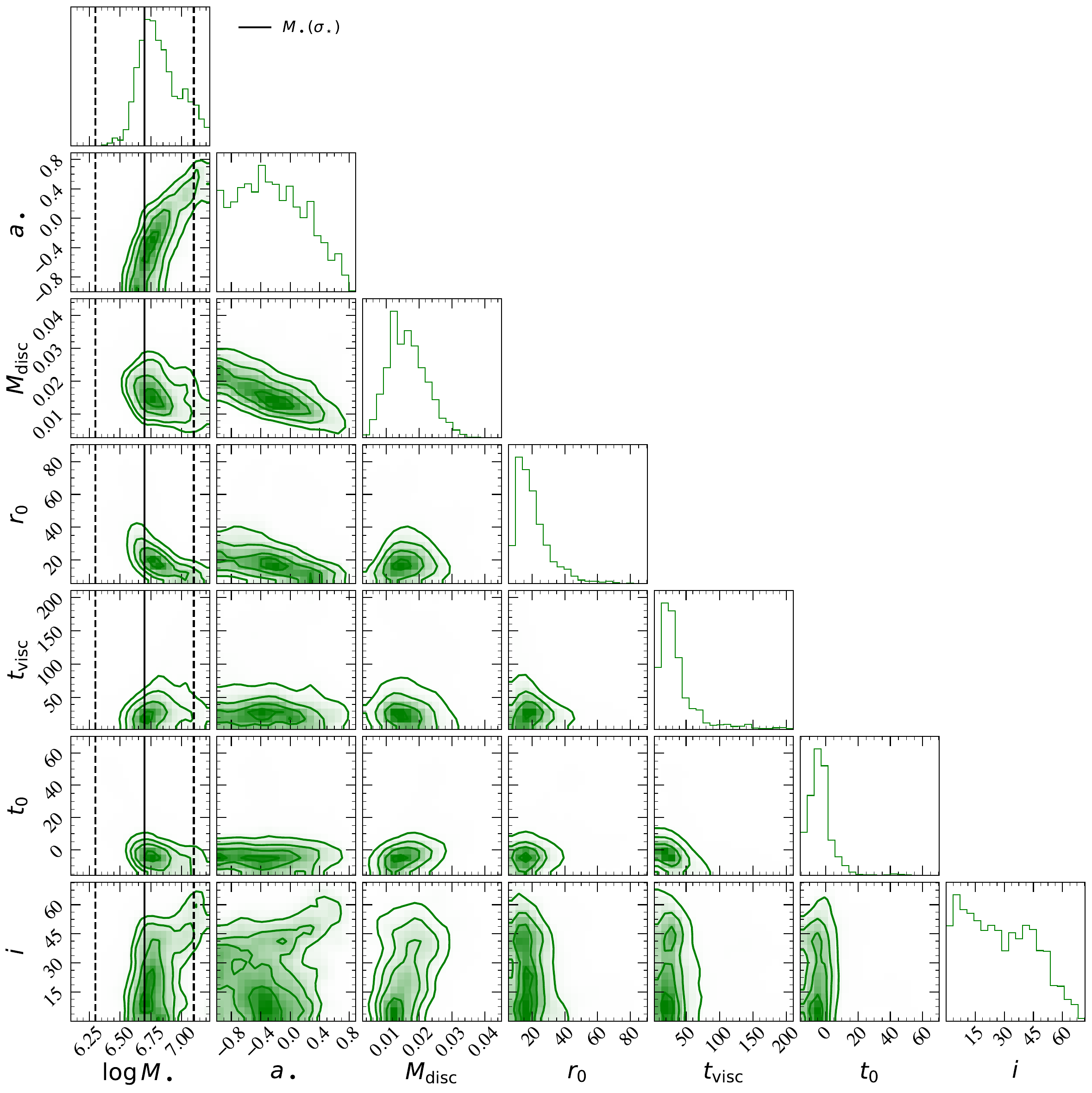}
    \caption{The posterior distributions of the disc parameters fit to the AT2019dsg light curves (Fig. \ref{fig:data_19dsg}). The parameters are presented in default \code\ units (see Table \ref{tab_pars}). The parameters are well constrained by the data, and the black hole mass is consistent with (but with smaller uncertainty than) the mass inferred from the velocity dispersion scaling relationship (black vertical lines denoting $1\sigma$ uncertainty on the velocity dispersion inference; left hand column).   }
    \label{fig:posterior-full}
\end{figure*}

The results of this fit are displayed in Figures \ref{fig:GRdisc_example} and \ref{fig:posterior-full}. In Figure \ref{fig:GRdisc_example} we show the model posterior light curves, along with the processed AT2019dsg data. Only the X-ray, UVW1 and $g$-band data and posteriors are shown for clarity, but all of the processed data from Figure \ref{fig:data_19dsg} is used for fitting purposes. The dotted curves show the disc contribution to the combined disc and early time decay model (shown by solid curves) in optical and UV frequencies. Shaded regions show the $10-90\%$ posterior range around the median (denoted by solid curves). The upper panel of Figure \ref{fig:GRdisc_example} shows that full dataset, while the lower panel shows the first 160 days to highlight the X-ray evolution. Clearly the model reproduces the gross long-term evolution of both the X-ray and optical-UV observations. 

The posterior distributions of the best-fitting parameters are displayed in Figure \ref{fig:posterior-full}. It is clear that the black hole mass, disc mass, initial radius, start time and viscous timescale are all well constrained from the data. The black hole spin and inclination are less well constrained, which is to be expected as the spin only effects the X-ray luminosity, of which there is less data. 

\begin{table}
    \centering
    \def\arraystretch{1.5}
    \begin{tabular}{c|c|c}
        Method & $\log_{10}M_\bullet/M_\odot$ & Reference \\
        \hline 
        Velocity dispersion & $6.7^{+0.4}_{-0.4}$ & \cite{Yao23} \\
        Plateau luminosity & $6.9^{+0.5}_{-0.4}$ & \cite{Mummery_et_al_2024} \\
        \code\ & $6.8_{-0.15}^{+0.2}$ & This work \\
        \hline 
    \end{tabular}
    \caption{Different methods of estimating the black hole mass of AT2019dsg, from either galactic properties, gross features of the late-time lightcurve, or a detailed light curve fit.  All are consistent, but a detailed light curve fit with \code\ produces the smallest uncertainty. }
    \label{tab:mass}
\end{table}

Importantly, the black hole mass implied by the \code\ model fits is consistent with, but with lower uncertainties than, the black hole mass which would be inferred from the black hole mass-velocity dispersion $(\sigma_\star)$ scaling relationship \citep{McConnellMa13, Greene20}, and the measured velocity dispersion of  AT2019dsg's host galaxy \citep{Yao23}. The $M_\bullet$ measurement inferred from $\sigma_\star$ is $\log_{10}M_\bullet/M_\odot = 6.7 \pm 0.4$, where the uncertainty is dominated by the intrinsic scatter in the \cite{Greene20} scaling relationship. The \code\ measurement is $\log_{10}M_\bullet/M_\odot = 6.8 \pm 0.2$, which is consistent with but more precise than the value found  from the plateau-mass scaling relationship developed in \cite{Mummery_et_al_2024} of $\log_{10}M_\bullet/M_\odot = 6.9\pm0.5$. These results are summarised in Table \ref{tab:mass}. 

Also well constrained is the initial radius of the matter ring, $r_0$. While there are systematic uncertainties in this quantity owing to the somewhat artificial nature of the initial condition, it is consistent with what would be expected for the circularisation radius of a stellar-type star around the best-fitting \code\ black hole mass. 

The circularisation radius is the radius at which  stellar debris are expected to return to post-disruption, and is roughly expected to correspond to the initial disc formation radius. A tidal disruption event occurs if the pericenter orbit of the stars incoming orbit is smaller than the tidal radius of the star, or in other words 
\begin{equation}
    r_p = {1\over \beta} r_T = {R_\star \over \beta} \left({M_\bullet \over M_\star}\right)^{1/3}, \quad \beta \geq 1. 
\end{equation}
The circularisation radius can then be found from conservation of angular momentum, and is given by $($assuming a solar-type star with mass radius relationship $R_\star \propto M_\star^{4/5})$
\begin{equation}\label{rcirc}
    \frac{R_{\rm circ}}{r_g} \simeq  \frac{2c^2R_{\odot}}{\beta G M_{\odot}} \left ( \frac{M_*}{M_{\odot}} \right)^{7/15} \left ( \frac{M_{\bullet}}{M_{\odot}} \right)^{-2/3}, 
\end{equation}
where the extra factor of 2 results from converting the angular momentum of a parabolic orbit into a circular orbit. For the best fitting AT2019dsg black hole mass $M_\bullet = 10^{6.8} M_\odot$ this radius is 
\begin{equation}
     \frac{R_{\rm circ}}{r_g} \approx { 27.6 \over \beta} \left ( \frac{M_*}{M_{\odot}} \right)^{7/15} ,
\end{equation}
close to the posterior values of $r_0$ (Figure \ref{fig:posterior-full}). Finally, the viscous timescale posterior tracks the decay timescale of the X-ray light curve (roughly $t_{\rm visc} \simeq 1$ month). Without X-ray information it is extremely difficult to constrain the evolutionary timescale of the disc, as the optical/UV flux from a TDE system in the disc dominated phase is nearly time-independent. 

\subsection{Physical information about the AT2019dsg system }
In addition to constraining the model parameters of the AT2019dsg, a full disc model like \code\ allows additional physical characteristics of the system to be probed. 
\begin{figure}
    \centering
    \includegraphics[width=\linewidth]{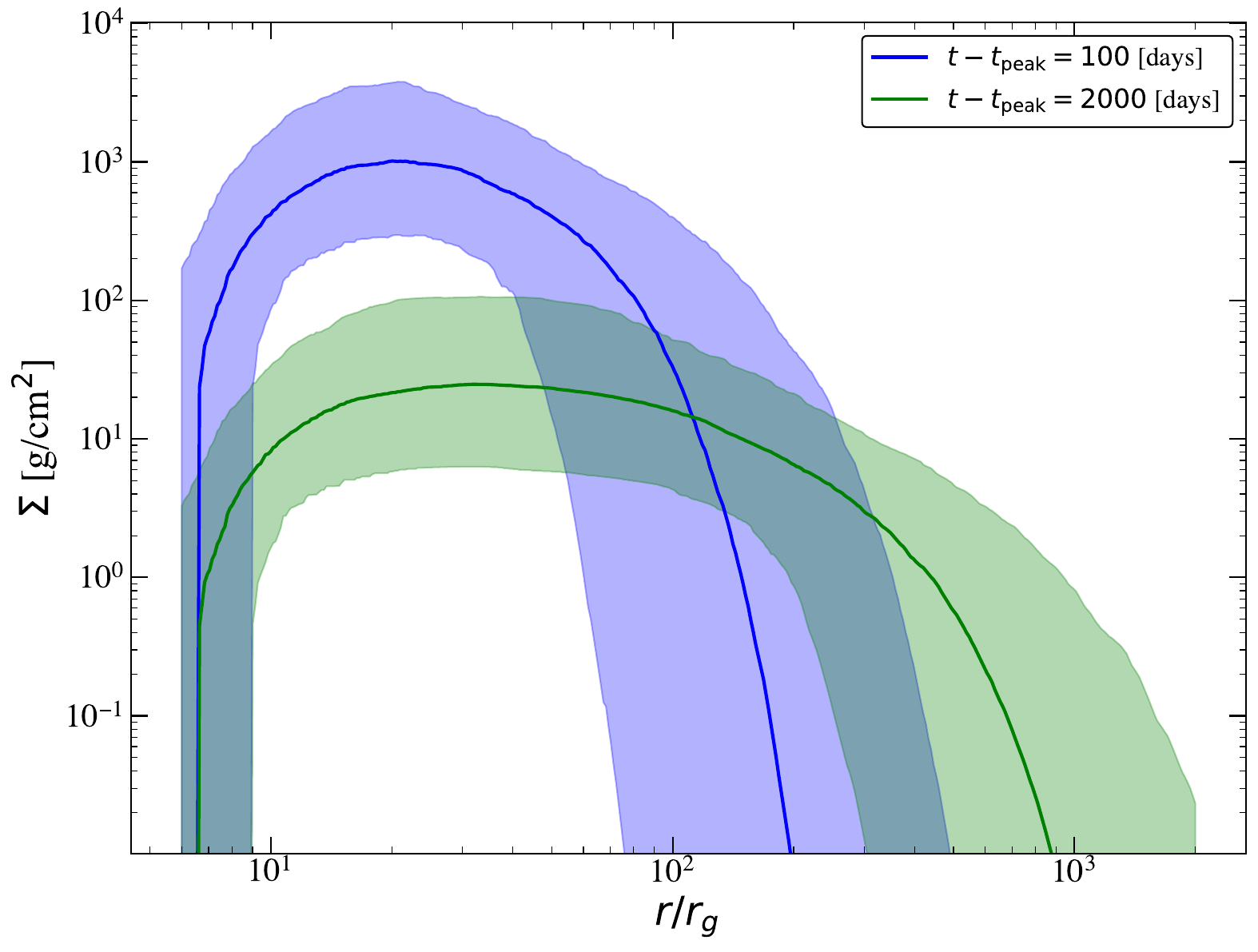}
    \includegraphics[width=\linewidth]{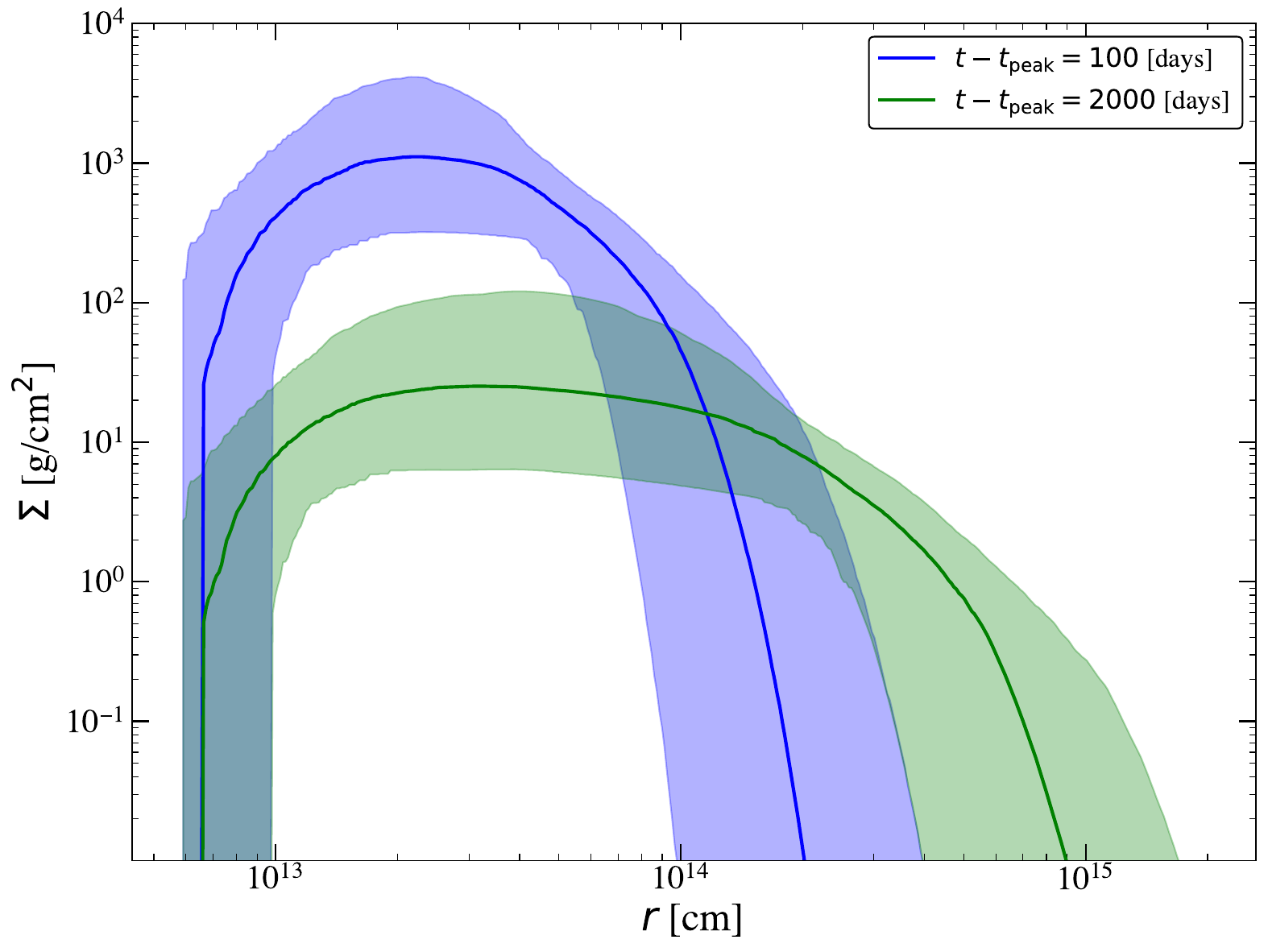}
    \caption{The $5-95\%$ contours of the disc surface density at an early epoch (blue) and a late-time epoch (green). The later time epoch can be differentiated by its lower amplitude. The posterior median is denoted by a solid curve at each epoch. The upper plot shows the density as a function of gravitational radii, while the lower plot shows the same contours in cgs units. Clearly, tight constraints can be placed on the size of the disc with \code\ models, which may be of use to future QPE science.  }
    \label{fig:density}
\end{figure}

\subsubsection{Disc density profiles}
For example, many models of the recently discovered class of X-ray transients known as quasi-periodic eruptions \citep[hereafter QPEs;][]{Miniutti2019,Giustini2020,Arcodia2021,Arcodia2024a,Guolo2024} suggest that the large-amplitude X-ray flares observed from these systems originate from the repeated crossing of a secondary object with an accretion flow surrounding a supermassive black hole \citep{Xian2021,Linial2023, Lu2023, Franchini2023}.  In some of these works, it has been suggested that this disc will, in many systems, have been seeded by a TDE \citep{Linial2023,Kaur2023}. To test these theories then it is essential to have an understanding of the physical size of the TDE disc as a function of time. This is possible with \code\ solutions. 

In Figure \ref{fig:density}, we display the surface density contours $(5-95\%)$ of the spreading disc solutions fit to AT2019dsg. The upper (blue) contours show the density contours at an early time ($t-t_0 = 100$ days), while the lower (green) contours show the density profile at a much later time ($t-t_0 = 2000$ days), corresponding to mid-October 2024. Clearly, an orbiting body at $r_{\rm orb} \simeq 3 \times 10^{14}$ cm would not have intercepted the disc at early times, but would at late times. This would correspond (for our best-fit black hole mass), to a QPE recurrence timescale of  $P_{\rm QPE} \simeq 6.5$ hours, where $P_{\rm QPE} = \pi \sqrt{r_{\rm orb}^3/GM_\bullet}$ (there are two flares per Keplarian orbit). 

As is clear in Figure \ref{fig:density}, in addition to spreading to larger radii the disc surface density drops with time, a result of continued accretion across the inner edge. The mass at later times can be simply estimated from the evolving outer edge of the disc as the disc's total angular momentum is formally conserved for a relativistic accretion flow with a vanishing ISCO stress. The angular momentum of the disc is dominated by the outer-most material, meaning that the quantity $M_{\rm disc}(t) \sqrt{r_{\rm out}(t)}$ remains $ \simeq {\rm constant}$ over the evolution of the flow. 

\begin{figure}
    \centering
    \includegraphics[width=\linewidth]{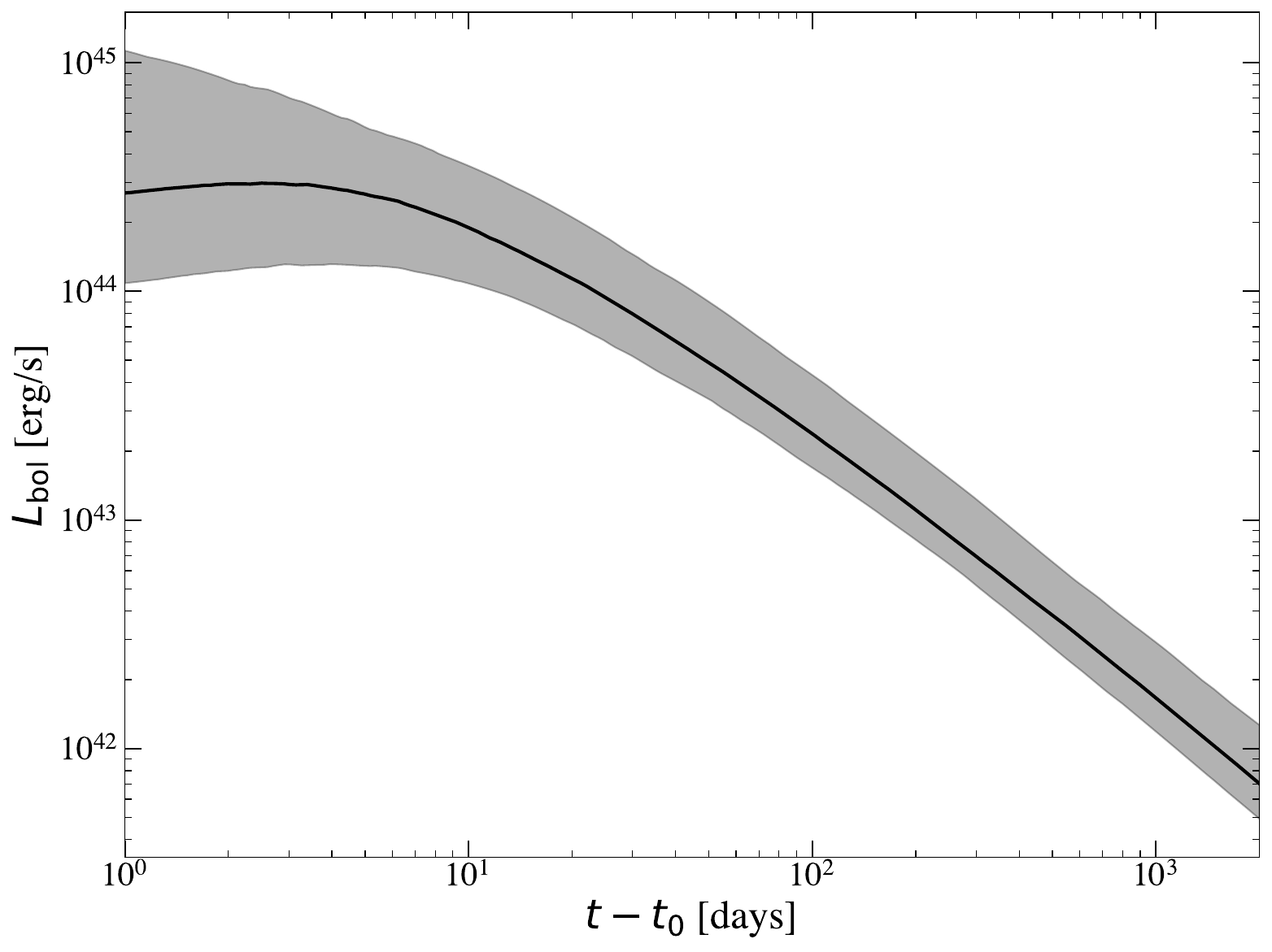}
    \includegraphics[width=\linewidth]{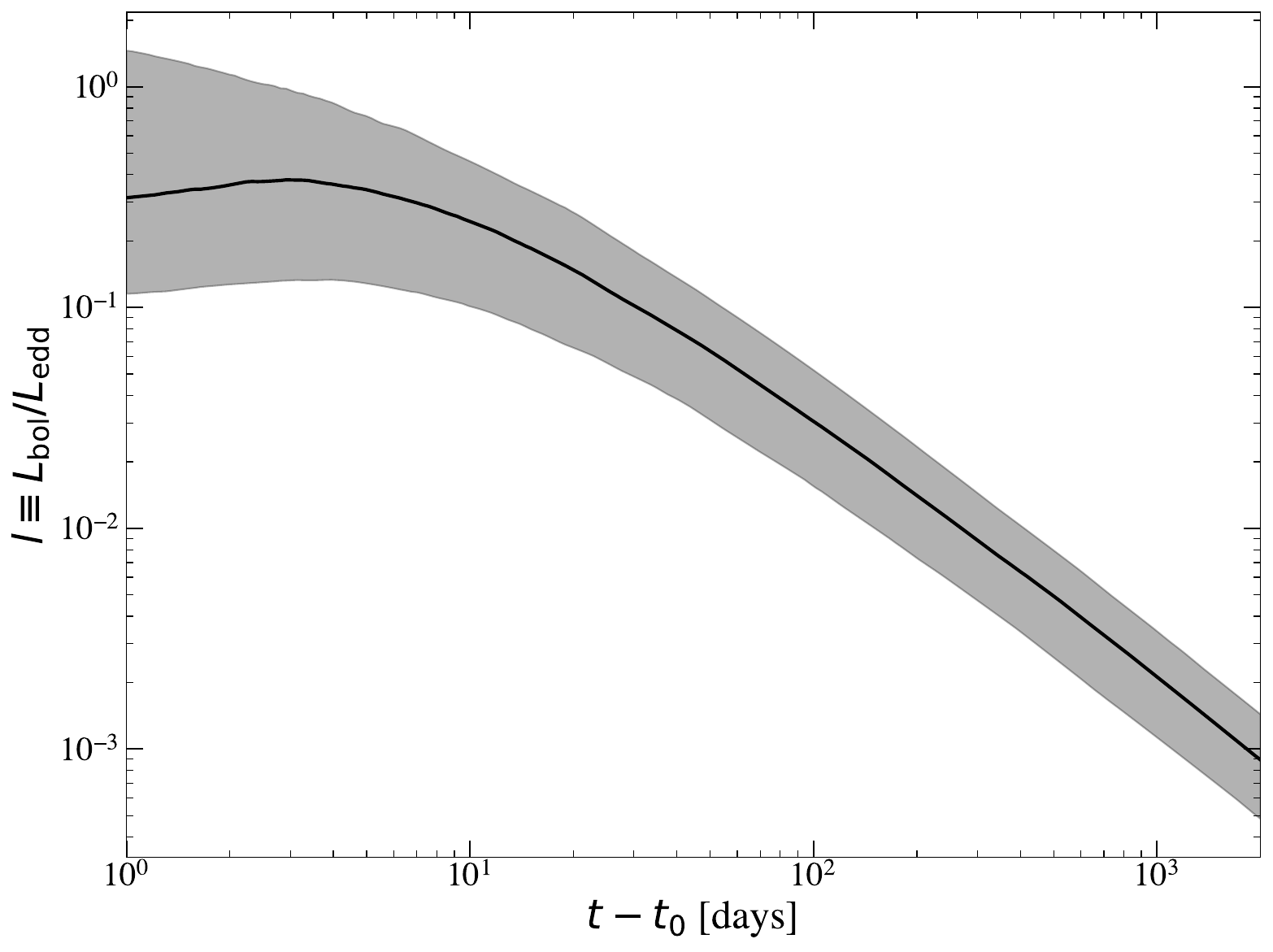}
    \includegraphics[width=\linewidth]{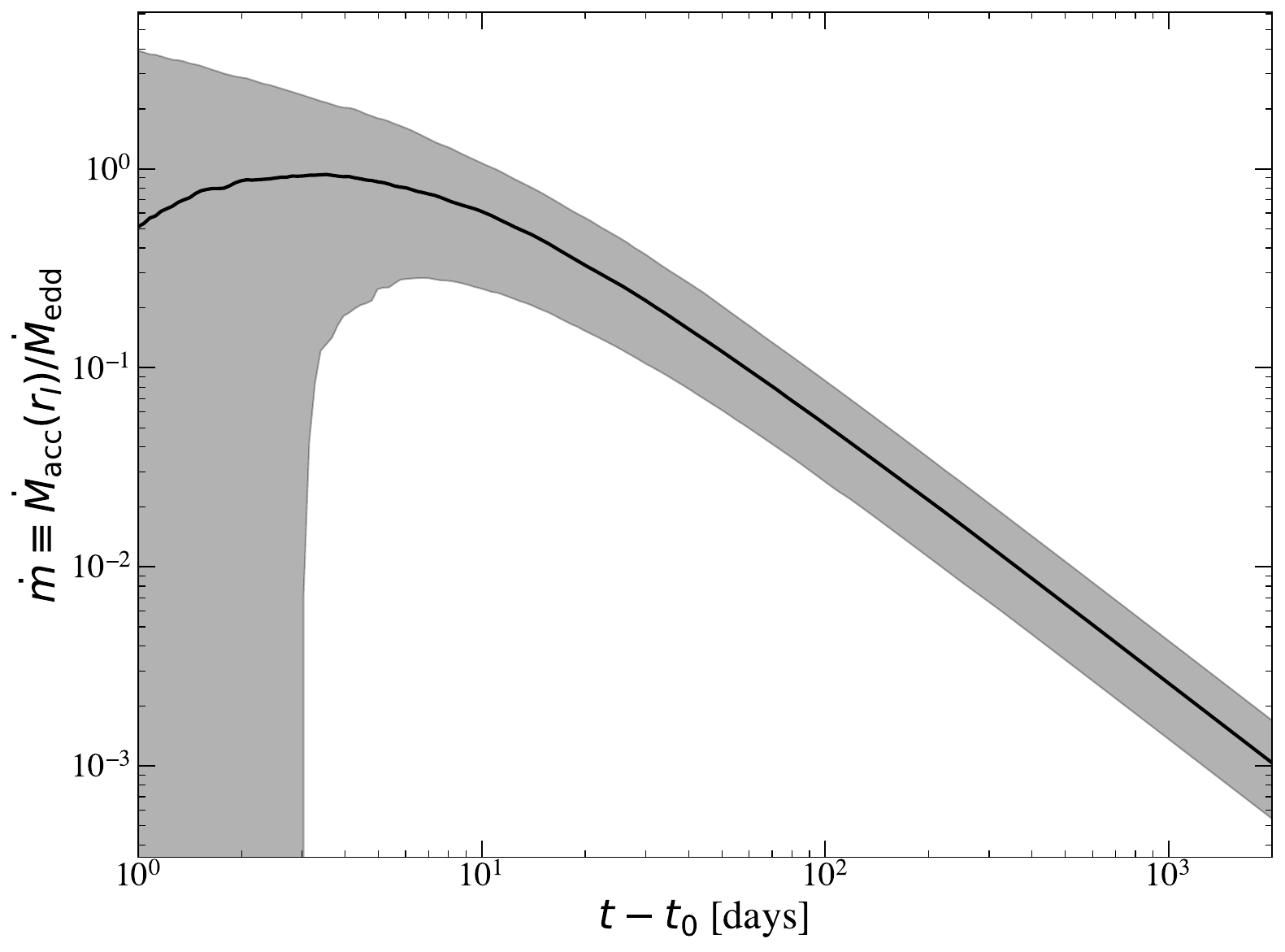}
    \caption{The posterior ($5-95\%$) bolometric luminosity (upper), and Eddington luminosity (middle) and ISCO accretion rate (lower) ratios for the AT2019dsg fit. Posterior medians are shown by solid curves. The Eddington ratio of the source is sub-Eddington (at all times when well constrained), as is required for the model to be valid. The Eddington ratio of AT2019dsg passes through $\sim 0.01$ at $t\sim 800$ days, coincident with the launching of an outflow detected in the radio. The bolometric correction for either optical or X-ray bands can be extremely large ($\sim 1000$) after $\sim 100$ days from peak.   }
    \label{fig:luminosity}
\end{figure}

\subsubsection{Bolometric disc properties}
In addition to the density profile of the accretion flow, other non-observable quantities of physical interest can be computed, including the bolometric luminosity, and the luminosity and accretion rate Eddington ratios of the system. 

In one sense this is a useful sanity check of the fit, as the model used is derived in the thin-disc limit which represent a series of approximations which will begin to break down as Eddington ratios significantly exceeding unity are reached. As can be seen in Figure \ref{fig:luminosity} the Eddington ratio of the AT2019dsg fit is robustly sub-Eddington at all times where the system is well constrained (typically times $> 3$ days into the evolution). This highlights that the thin-disc assumptions used are formally valid. 

On the other hand, these quantities are themselves interesting beyond merely acting as a sanity check of the model. Many properties of accreting disc systems, for example, are known to change at characteristic Eddington ratio's \citep[or at least they do in X-ray binary systems][]{Fender04}. 

This may be of particular interest for the study of the late-time radio re-brightening observed from many TDEs \citep[e.g.][]{Horesh21, Goodwin23}, an observed property which has been argued to be near-ubiquitous, with at least 40$\%$ of sources displaying this behaviour \citep{Cendes23}. It has been argued that this re-brightening may result from a late-time state transition within the disc \citep{Horesh21, Cendes23} when the Eddington ratio of the source passes through some characteristic value, typically $\sim 0.01$, such as is seen in many Galactic X-ray binaries \citep{Fender04}. 

AT2019dsg is one of these sources, showing a second radio re-brightening attributed to an outflow launched at $t \sim 800$ days post optical peak \citep{Cendes23}. It is extremely interesting to note that the Eddington ratio of the disc solutions derived here (measured both from the luminosity and ISCO accretion rate) cross the characteristic threshold of $\sim 1\%$ at precisely this time. 

Finally, it is interesting to contrast the disc's Bolometric luminosity with the observed luminosity scales of the system in the optical and X-ray bands. While at the earliest times the X-ray luminosity of the disc is within a factor $\sim 2-3$ of the disc bolometric luminosity, by day $\sim 100$ the X-ray luminosity is at least a factor $\sim 1000$ below the bolometric luminosity.   The reason for this is simple, and has been discussed extensively in \cite{MumBalb21a, Mummery_Wevers_23}: the X-ray luminosity of a TDE disc depends exponentially on the peak disc temperature, while the bolometric luminosity scales as $T_{\rm max}^4$. As the disc cools, the ``bolometric correction'' therefore grows exponentially, a natural resolution of the so-called ``missing-energy'' problem.




\section{Summary, discussion and conclusions}\label{summary}
\subsection{Discussion: potential model systematics}
As with any simplified model of a complex astrophysical system, a number of approximations have been employed in deriving this model, which could in principle lead to systematic errors in various situations. 

We reiterate the simplifications made to derive and solve the relativistic thin disc evolution equations. The disc is assumed to be 
\begin{enumerate}
    \item Thin,
    \item Evolving in the equatorial plane of the central black hole, 
    \item Accreting due to a local turbulent stress tensor,
    \item The time-evolution of the disc is smooth, 
    \item Radiating all of the locally liberated disc energy, 
    \item Radiates locally as a colour-corrected blackbody, 
    \item Has a vanishing ISCO stress. 
\end{enumerate}
All of these assumptions will be incorrect to some degree at some points in the evolution of the system, but most will likely become increasingly good approximations with time. Approximation (i) and (v) are really a statement regarding the Eddington ratio of the disc, which can be checked for each source post-fit (e.g., Figure \ref{fig:luminosity}). Approximation (ii) is statistically unlikely to be  true for any tidal disruption event, as there is no reason to suspect that the incoming stars orbit and the black hole's angular momentum will be aligned, but it is unclear how important this discrepancy will be for radii larger than $\sim 10r_g$. Approximation (ii) is likely to be a good approximation for X-ray binary systems, however. Assumption (iii) is common to all accretion models, and is an inevitable simplification required for the simple modelling of an inherently turbulent system. This turbulence will violate Assumption (iv) on timescale shorter than the disc viscous timescale, and short-timescale fluctuations in the disc lightcurves will not be captured by this model. This will likely lead to formally large discrepancies between the model and the data in X-ray light curves of tidal disruption events in particular, where turbulent disc fluctuations are exponentially enhanced \citep{MummeryTurner24}.  This short-timescale behaviour can be seen in Figure \ref{fig:GRdisc_example}, and should be mitigated by including systematic errors in the X-ray data of these sources. 

Assumption (vi) is supported by simulations of disc-atmosphere radiative transfer \citep[e.g.,][]{DavisHubeny06}, but still represents a simplification of the detailed physics of this process. It is known that uncertainties in $f_c$ typically dominate error budgets of black hole spin measurements \citep{Salvesen21}, which should caution against over-interpreting \code\ spin constraints. Assumption (vii) is employed for modelling simplicity, and is an extremely popular disc boundary condition. This assumption is known to be incorrect however, as has been shown in GRMHD simulations \citep[e.g.,][]{Noble10} and the detailed modelling of the spectra of X-ray binaries \citep{Mummery24Plunge}. One known systematic effect of making this assumption is that vanishing ISCO stress models systematically over-estimate black hole spins, owing to them compensating for the near-ISCO emission  of a finite ISCO stress system. 

In addition, as we have discussed throughout this paper, the initial condition employed by the model (that of an initial ring of material) is artificial, and should be mitigated with broad priors on the initial radius $r_0$ and start time $t_0$, parameters which should not be over-interpreted. 

Furthermore, the analytical solutions we employ in this paper are only themselves approximate at the ${\cal O}(1\%)$ level \citep{Mummery23a}, although this systematic uncertainty is likely significantly below other sources of error. 

Finally, further potential sources of spectral modification could in principle arise from any irradiation of the disc by an external source (e.g, a jet), or even self-irradiation by disc photons which have been gravitationally deflected to such a degree that they illuminate the disc. (This final effect becomes important if the angular momentum parameter of the black hole approaches unity.) The importance of these effects should be considered on a source-by-source basis, but appear likely to be sub-dominant. 

\subsection{Summary of AT2019dsg analysis and conclusions}
In this paper we have derived and presented the light curve and spectral fitting package \code. The purpose of this package is to provide constraints on the black hole and disc parameters of transient astronomical systems with spectra dominated by direct emission from an evolving accretion flow. 

\code\ self consistently produces disc light curves  by solving the time-dependent equations of disc mass, angular momentum and energy conservation, and then ray-tracing the resulting disc temperature profile. All relevant relativistic optics effects (Doppler and gravitational energy shifting, and gravitational lensing) are included, meaning that all leading order effects of general relativity are included. 

As an example of the possible uses of this package we have modelled the multi-band light curves of the tidal disruption event AT2019dsg. We have demonstrated that 
\begin{itemize}
    \item The \code\ model provides a good fit to the multi-band data
    \item The black hole mass inferred from the modelling is in good agreement with galactic scaling relationships, and estimates based on its broad light curve features 
    \item The thin disc assumptions employed in the modelling are valid at all times in the evolution of AT2019dsg
    \item The bolometric correction of the disc approaches extreme  values $\sim 100-1000$ after $\sim 100$ days from peak optical light
    \item The disc transitions through an Eddington ratio of $\sim 1\%$ at  a time $t\sim 800$ days after optical peak, coincident with the launching of a late-time radio flare 
    \item The AT2019dsg disc is now large enough for there to be detectable QPE's of periods of up to $\sim 10$ hours, provided there is an orbiting companion 
\end{itemize}
These latter three results highlight the importance of employing a full disc model, rather than fitting individual bands in isolation.  As only then can characteristics of the bolometric luminosity and disc density be constrained. These additional physical characteristics will likely have wide use in constraining the physics of evolving disc systems.

The primary usage of \code\ will likely be in analysing the forthcoming explosion in the number of tidal disruption events discovered by the Vera Rubin observatory as we enter the LSST era \citep{Bricman20}, although this code may be of use to the study of X-ray binaries. By providing constraints of the black hole masses of the tidal disruption events discovered in the coming decade, this package will enable independent probes of the black hole mass function at the lowest black hole masses, where there are currently the fewest observational 
constraints \citep{Greene20}.

\section*{Acknowledgments} 
AM is extremely grateful to Adelle Goodwin for stress testing an early version of \code\ and for numerous suggestions which improved the code. 

This work was supported by a Leverhulme Trust International Professorship grant [number LIP-202-014]. For the purpose of Open Access, AM has applied a CC BY public copyright licence to any Author Accepted Manuscript version arising from this submission. EN acknowledges support from NASA theory grant 80NSSC20K0540. AI acknowledges support from the Royal Society.

This research made use of the \texttt{Python} packages {\tt matplotlib} \citep{Hunter2007}, {\tt scipy} \cite{scipy}, {\tt numpy} \cite{numpy}, {\tt astropy} \cite{astropy},  {\tt emcee} \citep{EMCEE} and {\tt corner} \citep{corner}.

\section*{Data availability}
The \code\ package is available to download at the following repository:  \href{https://bitbucket.org/fittingtransientswithdiscs/fitted_public/src}{https://bitbucket.org/fittingtransientswithdiscs/fitted\_public/src}. 

The infrared, optical and UV data of all current tidal disruption events are publicly available at \href{https://github.com/sjoertvv/manyTDE}{https://github.com/sjoertvv/manyTDE}. All of the X-ray data used in this work can be downloaded from the HEASARC website \href{https://heasarc.gsfc.nasa.gov}{https://heasarc.gsfc.nasa.gov}.

\bibliographystyle{mnras}
\bibliography{andy}

\label{lastpage}
\end{document}